\documentclass[twocolumn,showpacs,aps,prd,floatfix]{revtex4}
\usepackage{graphicx}
\usepackage{dcolumn}
\usepackage{amsmath}
\usepackage{epsfig}
\usepackage{multirow}

\newcommand{\BABARPubYear} {09}
\newcommand{\BABARPubNumber} {040}
\newcommand{\SLACPubNumber} {14452}

\input{babarsym}
\newcommand{\pvec}{{\bf p}}
\newcommand{\half}{\mbox{$\frac{1}{2}$}}
\newcommand{\bei}{\begin{itemize}}
\newcommand{\eei}{\end{itemize}}
\newcommand{\beq}{\begin{equation}}
\newcommand{\eeq}{\end{equation}}
\newcommand{\beqn}{\begin{eqnarray}}
\newcommand{\eeqn}{\end{eqnarray}}
\newcommand{\beqns}{\begin{eqnarray*}}
\newcommand{\eeqns}{\end{eqnarray*}}

\newcommand{\equaref}[1]{Eq.~(\ref{eq:#1})}

\renewcommand{\secref}[1]{Sec.~\ref{sec:#1}}

\renewcommand{\figref}[1]{Fig.~\ref{fig:#1}}

\def\exp{{\rm exp}}

\def\min{{\rm min}}

\def\rPTbarkappa{\kern 0.18em\overline{\kern -0.18em r}{}^{\kappa}{}}
\def\rPTbarsigma{\kern 0.18em\overline{\kern -0.18em r}{}^{\sigma}{}}

\def\deltabarkappa{\kern 0.18em\overline{\kern -0.18em \delta}{}_r^{\kappa}}
\def\deltabarsigma{\kern 0.18em\overline{\kern -0.18em \delta}{}_r^{\sigma}}
\def\deltaTbarkappa{\kern 0.18em\overline{\kern -0.18em \delta}{}_T^{\kappa}}
\def\deltaTbarsigma{\kern 0.18em\overline{\kern -0.18em \delta}{}_T^{\sigma}}

\def\OC{X}
\def\OCbar{{\kern 0.18em\overline{\kern -0.18em \OC}}}

\def\BtoKpipi{\Bz\to K^+\pi^-\pi^0}
\def\Kpipi{K^+\pi^-\pi^0}

\def\deprime{\ensuremath{{\de^\prime}{}}}

\def\de{\DeltaE}

\def\Aqq{A_{q\bar q}}

\def\ABj{A_{B_j}}

\def\cat{c}

\def\a{\kappa}

\def\Amptpbar{\kern 0.18em\overline{\kern -0.18em {\cal A}}_{{\overline B^0} \rightarrow K^-\pi^+\pi^0}}

\def\Amptpbarkappa{\kern 0.18em\overline{\kern -0.18em A}{}^{\kappa}{}}
\def\Amptpbarsigma{\kern 0.18em\overline{\kern -0.18em A}{}^{\sigma}{}}
\def\AmpAll{|{\cal A}|^2+|\overline{\cal A}|^2}

\def\Tbarkappa{\kern 0.18em\overline{\kern -0.18em T}{}^{\kappa}{}}
\def\Tbarsigma{\kern 0.18em\overline{\kern -0.18em T}{}^{\sigma}{}}
\def\Pbarkappa{\kern 0.18em\overline{\kern -0.18em P}{}^{\kappa}{}}
\def\Pbarsigma{\kern 0.18em\overline{\kern -0.18em P}{}^{\sigma}{}}

\def\CP{{\em CP}}

\def\Nbpm{{\kern 0.18em\overline{\kern -0.18em N}}^{+-}}
\def\Nbmp{{\kern 0.18em\overline{\kern -0.18em N}}^{-+}}

\def\Mu{\mu}
\def\Chi2MinaMu{\chi^2_{\min ;\a,\Mu}}
\def\Chi2MinMu{\chi^2_{\min ;\Mu}(a)}

\def\TM{{\rm TM}}
\def\SCF{{\rm SCF}}

\def\fscfave{\langle f_{\rm SCF} \rangle_{\rm{DP}}}
\def\fscf{f_{\rm SCF}}

\def\abar{\bar{a}}

\def\Bbar{\kern 0.18em\overline{\kern -0.18em B}{}\xspace}

\def\BRpmb{{\cal \kern 0.18em\overline{\kern -0.18em  B}}{}_{\rho\pi}^{+-}}
\def\BRmpb{{\cal \kern 0.18em\overline{\kern -0.18em  B}}{}_{\rho\pi}^{-+}}

\def\BRipmb{{\cal \kern 0.18em\overline{\kern -0.18em  B}}{}_{\rho^+\pi^-}}
\def\BRimpb{{\cal \kern 0.18em\overline{\kern -0.18em  B}}{}_{\rho^-\pi^+}}

\def\Abar{\kern 0.18em\overline{\kern -0.18em A}{}}
\def\abar{\kern 0.18em\overline{\kern -0.18em a}{}}

\def\ie{{\em i.e.}}

\newcommand{\optbar}[1]{\shortstack{{\tiny(\rule[.4ex]{.8em}{.2mm})}\\[-.7ex]$#1$}}
\def\AorAbar    {\kern 0.18em\optbar{\kern -0.18em {\cal A}}{}\xspace}
\def\aorabar    {\kern 0.18em\optbar{\kern -0.18em {a}}{}\xspace}
\def\sorsbar    {\kern 0.18em\optbar{\kern -0.18em {s}}{}\xspace}
\def\corcbar    {\kern 0.18em\optbar{\kern -0.18em {c}}{}\xspace}
\def\PhiorPhibar    {\kern 0.18em\optbar{\kern -0.18em {\Phi}}{}\xspace}
\def\phiorphibar    {\kern 0.18em\optbar{\kern -0.18em {\phi}}{}\xspace}
\def\deltaordeltabar    {\kern 0.18em\optbar{\kern -0.18em {\delta}}{}\xspace}

\begin{document}

\begin{flushleft}
SLAC-PUB-\SLACPubNumber \\
\babar-PUB-\BABARPubYear/\BABARPubNumber
\end{flushleft}

\title{\Large \bf\boldmath Amplitude Analysis of $\Bz\to\Kp\pim\piz$ and Evidence of Direct \CP~Violation in~$\B\to K^*\pi$ decays}
\bigskip

%
\author{J.~P.~Lees}
\author{V.~Poireau}
\author{E.~Prencipe}
\author{V.~Tisserand}
\affiliation{Laboratoire d'Annecy-le-Vieux de Physique des Particules (LAPP), Universit\'e de Savoie, CNRS/IN2P3,  F-74941 Annecy-Le-Vieux, France}
\author{J.~Garra~Tico}
\author{E.~Grauges}
\affiliation{Universitat de Barcelona, Facultat de Fisica, Departament ECM, E-08028 Barcelona, Spain }
\author{M.~Martinelli$^{ab}$}
\author{D.~A.~Milanes$^{a}$}
\author{A.~Palano$^{ab}$ }
\author{M.~Pappagallo$^{ab}$ }
\affiliation{INFN Sezione di Bari$^{a}$; Dipartimento di Fisica, Universit\`a di Bari$^{b}$, I-70126 Bari, Italy }
% ...
\collaboration{The \babar\ Collaboration}
\noaffiliation

\date{\today}

\begin{abstract}
\noindent We analyze the decay $\Bz\to\Kp\pim\piz$ with a sample of 454 million $\B\Bbar$ events collected by the \babar\ detector at the \pep2\ asymmetric-energy \B~factory at SLAC, and extract the complex amplitudes of seven interfering resonances over the Dalitz plot. These results are combined with amplitudes measured in $\Bz\to\KS\pi^+\pim$ decays to construct isospin amplitudes from $\Bz\to\Kstar\pi$ and $\Bz\to\rho K$ decays. We measure the phase of the isospin amplitude $\Phi_{3\over2}$, useful in constraining the CKM unitarity triangle angle $\gamma$ and evaluate a \CP~rate asymmetry sum rule sensitive to the presence of new physics operators. We measure direct \CP~violation in $\Bz\to\Kstarp\pim$ decays at the level of $3~\sigma$ when measurements from both $\Bz\to\Kp\pim\piz$ and $\Bz\to\KS\pi^+\pim$ decays are combined. 
\end{abstract}


\maketitle

\section{INTRODUCTION}
\label{sec:Introduction}
In the Standard Model (SM), \CP~violation in weak interactions is parametrized by an irreducible complex phase in the Cabibbo-Kobayashi-Maskawa (CKM) quark mixing matrix~\cite{Cabibbo,Kobayashi}. The unitarity of the CKM matrix is typically expressed as a triangular relationship among its parameters such that decay amplitudes are sensitive to the angles of the triangle denoted $\alpha,\beta,\gamma$. Redundant measurements of the parameters of the CKM matrix provide an important test of the SM, since violation of the unitarity condition would be a signature of new physics. The angle $\gamma$ remains the least well measured of the CKM angles. Tree amplitudes in $\B\to K^*\pi$ decays are sensitive to $\gamma$ but are Cabibbo-suppressed relative to loop-order (penguin) contributions involving radiation of either a gluon (QCD penguins) or a photon (electroweak penguins or EWPs) from the loop. 

It has been shown that QCD penguin contributions can be eliminated by constructing a linear combination of $\Bz\to\Kstarp\pim$ and $\Bz\to\Kstarz\piz$ weak decay amplitudes that is pure (isospin) $I = {3\over2}$~\cite{Ciuchini:2006kv}, 

\begin{equation}
\label{eq:A32}
{\cal A}_{3 \over 2}(\Kstar\pi) = {1 \over \sqrt{2}} {\cal A}(\Bz\to\Kstarp\pim) + {\cal A}(\Bz\to\Kstarz\piz).
\end{equation}

\noindent Since a transition from $I = {1\over2}$ to $I = {3\over2}$ is possible only via $\Delta I = 1$ operators, ${\cal A}_{3\over2}$ must be free of $\Delta I = 0$, namely QCD contributions. The weak phase of ${\cal A}_{3\over2}$, given by $\Phi_{3\over2} = -{1\over2}\rm{Arg}\big({\overline{\cal A}_{3\over2} / {\cal A}_{3\over2}}\big)$, is equal to the CKM angle $\gamma$ in the absence of EWP operators~\cite{Gronau:2006qn}. Here, $\overline{\cal A}_{3\over2}$ denotes the \CP~conjugate of the amplitude in~\equaref{A32}.

The relative magnitudes and phases of the $\Bz\to\Kstarp\pim$ and $\Bz\to\Kstarz\piz$ amplitudes in~\equaref{A32} are measured from their interference over the available decay phase space (Dalitz plot or DP) to the common final state $\Bz\to\Kp\pim\piz$. The phase difference between $\Bz\to\Kstarp\pim$ and $\Bzb\to\Kstarm\pip$ is measured in the DP analysis of the self-conjugate final state~$\Bz\to \KS\pi^+\pim$~\cite{babar-kspipi} where the strong phases cancel. This argument is extended to $\Bz\to\rho K$ decay amplitudes~\cite{Wagner,Antonelli2010197} where an isospin decomposition of amplitudes gives

\begin{equation}
\label{eq:A32_rhoK}
{\cal A}_{3 \over 2}(\rho K) = {1 \over \sqrt{2}} {\cal A}(\Bz\to\rho^-\Kp) + {\cal A}(\Bz\to\rho^0\Kz).
\end{equation}

\noindent Here, the $\Bz\to\rho^-\Kp$ and $\Bz\to\rho^0\Kz$ decays do not decay to a common final state preventing a direct measurement of their relative phase. The amplitudes in~\equaref{A32_rhoK} do, however, interfere with the $\Bz\to\Kstarp\pim$ amplitude in their decays to $\Bz\to\Kp\pim\piz$ and $\Bz\to\KS\pi^+\pim$ final states so that an indirect measurement of their relative phase is possible. 

The \CP~rate asymmetries of the isospin amplitudes ${\cal A}_{3 \over 2}(\Kstar\pi)$ and ${\cal A}_{3 \over 2}(\rho K)$ have been shown to obey a sum rule~\cite{Gronau:sumrule}, 

\begin{equation}
\label{eq:sumrule}
|\overline{{\cal A}}_{3 \over 2}(\Kstar\pi)|^2 - |{\cal A}_{3 \over 2}(\Kstar\pi)|^2 = |{\cal A}_{3 \over 2}(\rho K)|^2 - |\overline{{\cal A}}_{3 \over 2}(\rho K)|^2.
\end{equation} 

\noindent This sum rule is exact in the limit of SU(3) symmetry and large deviations could be an indication of new strangeness violating operators. Measurements of $\Bz\to\Kstar\pi$ and $\Bz\to\rho K$ amplitudes are used to evaluate~\equaref{sumrule}.

We present an update of the DP analysis of the flavor-specific $\Bz\to\Kp\pim\piz$ decay from Ref.~\cite{babar-kpipi} with a sample of 454 million $\B\Bbar$ events. The isobar model used to parametrize the complex amplitudes of the intermediate resonances contributing to the final state is presented in~\secref{DecayAmplitudes}. The~\babar~detector and data set are briefly described in~\secref{DetectorAndData}. The efficient selection of signal candidates is described in~\secref{selection} and the unbinned maximum likelihood (ML) fit performed with the selected events is presented in~\secref{ML}. The complex amplitudes of the intermediate resonances contributing to the $\Bz\to\Kp\pim\piz$ decay are extracted from the result of the ML fit in~\secref{Results} together with the accounting of the systematic uncertainties in~\secref{Systematics}. Several important results are discussed in~\secref{Interpretation}. Measurements of $\Bz\to\rho K$ from this article and Ref.~\cite{babar-kspipi} are used to produce a measurement of $\Phi_{3\over2}$ using~\equaref{A32_rhoK}. It is shown that the large phase difference between $\Bz\to\Kstarp(892)\pim$ and $\Bz\to\Kstarz(892)\piz$ amplitudes makes a similar measurement using~\equaref{A32} impossible with the available data set. We find that the sum rule in~\equaref{sumrule} holds within the experimental uncertainty. Additionally, we find evidence for a direct \CP~asymmetry in $\Bz\to\Kstarp\pim$ decays when the results of Ref.~\cite{babar-kspipi} are combined with measurements in this article. The conventions and results of Ref.~\cite{babar-kspipi} are summarized where necessary. Finally in~\secref{Summary}, we summarize our results.
\section{Analysis Overview}
We present a DP analysis of the $\Bz\to\Kp\pim\piz$ decay in which we measure the magnitudes and relative phases of five resonant amplitudes: $\rho(770)^-\Kp$, $\rho(1450)^-\Kp$, $\rho(1700)^-\Kp$, $\Kstar(892)^+\pim$, $\Kstar(892)^0\piz$, two $K\pi$ S-waves: $(K\pi)_0^{*0},~(K\pi)_0^{*+}$, and a non-resonant (NR) contribution, allowing for \CP~violation. The notation for the S-waves denotes phenomenological amplitudes described by coherent superpositions of an elastic effective range term and the $\Kstar_0(1430)$ resonances~\cite{babar-latham}. Here, we describe the decay amplitude formalism and conventions used in this analysis. 

\label{sec:DecayAmplitudes}
The $\Bz\to\Kp\pim\piz$ decay amplitude is a function of two independent kinematic variables: we use the squares of the invariant masses of the pairs of particles $\Kp\pim$ and $\Kp\piz$, $x=m_{\Kp\pim}^2$ and $y=m_{\Kp\piz}^2$. The total decay amplitude is a linear combination of $k$ isobars, each having amplitude ${\cal A}_k$ given by:

\begin{equation}
\label{eq:isobars}
\AorAbar_k =  \aorabar_ke^{i\PhiorPhibar_k} \int_{\rm{DP}}  f_k(J,x,y)~dx\ dy
\end{equation}

\noindent where 

\begin{equation}
\label{eq:lsnorm}
\left|\int_{\rm{DP}}  f_k(J,x,y)~dx\ dy\right| = 1. 
\end{equation}

\noindent Here, $\overline{\cal A}_k$ denotes the \CP~conjugate amplitude, and $\aorabar_ke^{i\PhiorPhibar_k}$ is the complex coefficient of the isobar. The normalized decay dynamics of the intermediate state are specified by the functions $f_k$ that for a spin-$J$ resonance in the $\Kp\pim$ decay channel describe the angular dependence $T_k(J,x,y)$, Blatt-Weisskopf centrifugal barrier factor~\cite{BlattWeissk} $B_k(J,x)$, and mass distribution of the resonance $L_k(J,x)$: 

\begin{equation}
\label{eq:fequalRT}
f_k(J,x,y) =  T_k(J,x,y) \times B_k(J,x) \times L_k(J,x).
\end{equation}

\noindent The branching fractions ${\cal B}_k$ (\CP~averaged over $\Bz$ and $\Bzb$), and \CP~asymmetry, $A_{C\!P}(k)$ are given by:

\begin{eqnarray} 
\label{eq:PartialFractions}
  {\cal B}_k &=& \frac{\left|{\cal A}_k\right|^2 + \left|\overline{\cal A}_k\right|^2}{\left|\sum\limits_{j}{\cal A}_j\right|^2 + \left|\sum\limits_{j}\overline{\cal A}_j\right|^2} \times \frac{N_{\rm{sig}}}{N_{\BB}\big<\epsilon\big>_{\rm{DP}}},\\
  A_{C\!P}(k) &=& {\frac{|\overline{\cal A}_k|^2 - |{\cal A}_k|^2} {|\overline{\cal A}_k|^2 + |{\cal A}_k|^2}} = \frac{\overline{a}_k^2 - a_k^2} {\overline{a}_k^2 + a_k^2}.
\end{eqnarray}

\noindent where $N_{\rm sig}$ is the number of $\Bz\to\Kp\pim\piz$ events selected from a sample of $N_{\BB}$ \B-meson decays. The average DP efficiency, $\big<\epsilon\big>_{\rm DP}$ is given by

\begin{equation}
\label{eq:Daleff}
\frac{\left|\sum\limits_{k} (a_ke^{i\Phi_k} + \overline{a}_ke^{i\overline{\Phi}_k})\int_{\rm{DP}}\epsilon(x,y) f_k(J,x,y)~dx\ dy\right|}{\left|\sum\limits_{k} (a_ke^{i\Phi_k} + \overline{a}_ke^{i\overline{\Phi}_k})\int_{\rm{DP}}f_k(J,x,y)~dx\ dy\right|},
\end{equation}

\noindent where $\epsilon(x,y)$ is the DP dependent signal selection efficiency. 

\begin{figure}
  \epsfig{file=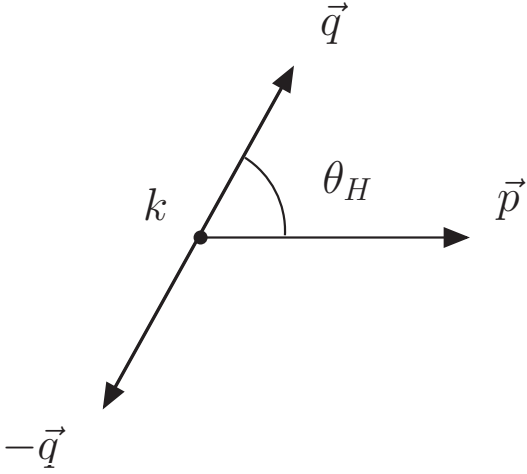,width=8cm}
  \caption {\label{fig:Helicity} The helicity angle ($\theta_H$) and momenta of particles ($\vec{q},\vec{p}$) in the rest frame of a resonance $k$.}
\end{figure}

We use the Zemach tensor formalism~\cite{Zemach} for the angular distribution $T(J,x,y)$ of a process by which a pseudoscalar $\B$-meson produces a spin-$J$ resonance in association with a bachelor pseudoscalar meson. We define $\vec{p}$ and $\vec{q}$ as the momentum vectors of the bachelor particle and resonance decay product, respectively, in the rest frame of the resonance $k$. The choice of the resonance decay product defines the helicity convention for each resonance where the cosine of the helicity angle is $\cos{\theta_H} = \vec{p}\cdot\vec{q} / (|\vec{p}||\vec{q}|)$. We choose the resonance decay product with momentum $\vec{q}$ to be the $\pim$ for $\Kp\pim$ resonances, the $\piz$ for $\pim\piz$ resonances, and the $\Kp$ for $\piz\Kp$ resonances (see~\figref{Helicity}).

The decay of a spin-$J$ resonance into two pseudoscalars is damped by a Blatt-Weisskopf barrier factor, characterized by the phenomenological radius $R$ of the resonance. The Blatt-Weisskopf barrier factors $B(J,x)$ are normalized to $1$ when $\sqrt{x}$ equals the pole mass $M$ of the resonance. We parametrize the barrier factors in terms of $z = |\vec{q}|R$ and $z_0 = |\vec{q}_0|R$ where $|\vec{q}_0|$ is the value of $|\vec{q}|$ when $\sqrt{x} = M$. The angular distributions and Blatt-Weisskopf barrier factors for the resonance spins used in this analysis are summarized in~\tabref{Zblatt}. 

\begin{table}[h]
\begin{center}
\caption{\label{tab:Zblatt} The angular distributions $T(J,x,y)$, and Blatt-Weisskopf barrier factors $B(J,x)$, for a resonance of spin-$J$ decaying to two pseudoscalar mesons. } 
\begin{tabular}{ccc} \hline\hline
Spin-$J$ & $T(J,x,y)$ & $B(J,x)$ \\ \hline
&& \\
0 &           $\displaystyle{1}$                                              &              $\displaystyle{1}$     \\
&& \\
1 & $\displaystyle{-2|\vec{p}||\vec{q}|\cos{\theta_H}}$                       &    $\displaystyle{\sqrt{{1 + z^2_0}\over{1 + z^2}}}$  \\
&& \\
2 & $\displaystyle{{4\over3}|\vec{p}|^2|\vec{q}|^2(3\cos^2{\theta_H} - 1)}$   &  $\displaystyle{\sqrt{{9 + 3z^2_0 + z^4_0}\over{9 + 3z^2 + z^4}}}$ \\
&& \\
\hline\hline
\end{tabular}
\end{center}
\end{table}

We use the relativistic Breit-Wigner~(RBW) lineshape to describe the $K^*(892)^{+,0}$ resonances,

\begin{equation}
\label{eq:nominalBW}
L^{\rm{RBW}}(J,x) = \frac{1}{M^2 - x - i M\Gamma(J,x)}.
\end{equation}

\noindent Here, the mass-dependent width $\Gamma(J,x)$ is defined by 

\begin{equation}
\label{eq:s-dependentWidth}
\Gamma(J,x) = \Gamma^0\frac{M}{\sqrt{x}}\left(\frac{|\vec{q}|}{|\vec{q}_0|}\right)^{\!2J+1}B(J,x)^2,
\end{equation}

\noindent where $\Gamma^0$ is the natural width of the resonance. 

The Gounaris-Sakurai~(GS) parametrization~\cite{Gounaris:1968mw} is used to describe the lineshape of the broad $\rho(770)^-$, $\rho(1450)^-$ and $\rho(1700)^-$ resonances decaying to two pions:

\begin{equation}
\label{eq:GS}
L^{\rm{GS}}(J,x) = \frac{1+d\Gamma^0/M}{M^2+g(x)-x -i M\Gamma(J,x)},
\end{equation}

\noindent where $\Gamma(J,x)$ is defined in~\equaref{s-dependentWidth}. Expressions for the constant $d$ and the function $g(x)$ in terms of $M$ and $\Gamma^0$ are given in Ref.~\cite{Gounaris:1968mw}. The parameters of the $\rho$ lineshapes, $M$, and $\Gamma^0$ are taken from Ref.~\cite{babar-rhopi} using updated lineshape fits with data from $\ep e^-$ annihilation~\cite{CMD2pipi} and $\tau$ decays~\cite{aleph-tau}. 

An effective-range parametrization was suggested~\cite{Estabrooks} for the $K\pi$ scalar amplitudes, $(K\pi)^{*+}_0$ and $(K\pi)^{*0}_0$ which dominate for $m_{K\pi}<2~\gevcc$, to describe the slowly increasing phase as a function of the $K\pi$ mass. We use the parametrization chosen in the LASS experiment~\cite{LASS}, tuned for \B-meson decays~\cite{babar-latham}:

\begin{eqnarray}
\label{eq:LASS}
L^{\rm{LASS}} (x) &=& \frac{\sqrt{x}/M^2}{|\vec{q}| \cot{\delta_B} - i|\vec{q}|} \nonumber \\
                & & + e^{2i \delta_B} \frac{\Gamma^0/|\vec{q}_0|}{M^2 - x - i M\Gamma(0,x)}, 
\end{eqnarray}

\noindent with

\begin{equation}
  \label{eq:LASSphase}
  \cot{\delta_B} = \frac{1}{a |\vec{q}|} + \frac{1}{2}\,r\,|\vec{q}|\,,
\end{equation}

\noindent where $a$ is the scattering length, and $r$ the effective range (see~\tabref{nominal}). We impose a cutoff for the $K\pi$ S-waves so that $L^{\rm{LASS}}$ is given only by the second term in~\equaref{LASS} for $\sqrt{x}>1.8~\gevcc$. Finally, the NR $\Kp\pim\piz$ amplitude is taken to be constant across the DP. 

In addition to the seven resonant amplitudes and the NR component described above we model the contributions to the $\Kp\pim\piz$ final state from $\Bz\to\overline{D}^0\piz$ and $\Bz\to\Dm\Kp$ with a double Gaussian distribution given by 

\begin{eqnarray}
  \label{eq:DG}
  L^{\rm{DG}}(x) &=& \frac{1-f}{\sigma_1}\exp\left[-\frac{(M_1-\sqrt{x})^2}{2\sigma_1^2}\right] \nonumber\\
                & & + \frac{f}{\sigma_2}\exp\left[-\frac{(M_2-\sqrt{x})^2}{2\sigma_2^2}\right]. 
\end{eqnarray}

\noindent The fraction $f$ is the relative weight of the two Gaussian distributions parametrized by the masses $M_1,~M_2$ and widths $\sigma_1,~\sigma_2$. The $D$ mesons are modeled as non-interfering isobars and are distinct from the charmless signal events.

\begin{table}[h]
\begin{center}
\caption{\label{tab:nominal}The model for the $\Bz\to\Kp\pim\piz$ decay comprises a non-resonant (NR) amplitude and seven intermediate states listed below. The three lineshapes are described in the text and the citations reference the parameters used in the fit. We use the same LASS parameters~\cite{LASS} for both neutral and charged $K\pi$ systems.}
\begin{tabular}{ccl} \hline\hline
Resonance & Lineshape & Parameters\\ \hline
\multicolumn{3}{c}{{Spin $J=1$}}\\ 
&& \\
$\rho(770)^-$              & GS~\cite{babar-rhopi}  & $M=775.5\pm 0.6~\mevcc$\\                                                           
                           &     & $\Gamma^0=148.2\pm 0.8~\mev$ \\ 
                           &     & $R=0~(\gevc)^{-1}$\\
$\rho(1450)^-$             & GS~\cite{babar-rhopi}  & $M=1409\pm 12~\mevcc$\\
                           &     & $\Gamma^0=500\pm 37~\mev$\\
                           &     & $R=0~(\gevc)^{-1}$\\
$\rho(1700)^-$             & GS~\cite{babar-rhopi}   & $M=1749\pm 20~\mevcc$\\ 
                           &     & $\Gamma^0=235\pm 60~\mev$\\
                           &     & $R=0~(\gevc)^{-1}$\\
&& \\
$\Kstar(892)^+$             & RBW~\cite{PDG}  & $M=891.6\pm 0.26~\mevcc$\\
                           &      & $\Gamma^0=50\pm 0.9~\mev$\\
                           &      & $R=3.4~(\gevc)^{-1}$\\
$\Kstar(892)^0$             & RBW~\cite{PDG}  & $M=896.1\pm 0.27~\mevcc$\\
                           &      & $\Gamma^0=50.5\pm 0.6~\mev$\\
                           &      & $R=3.4~(\gevc)^{-1}$\\
\multicolumn{3}{c}{{Spin $J=0$}}\\ 
&& \\
$(K\pi)^{*+}_0$, $(K\pi)^{*0}_0$  & LASS~\cite{LASS} & $M=1412\pm 3~\mevcc$\\
                                &      & $\Gamma^0=294\pm 6~\mev$\\
                                &      & $a=2.07\pm 0.10~(\gevc)^{-1}$\\ 
                                &      & $r=3.32\pm 0.34~(\gevc)^{-1}$\\ 
&& \\
NR                         & Constant &  \\ \hline
\multicolumn{3}{c}{Non-interfering components}                              \\ 
&& \\
$\overline{D}^0$ &  DG                          & $M_1=1862.3~\mevcc$\\
                 &                              & $\sigma_1=7.1~\mev$\\
                 &                              & $M_2=1860.1~\mevcc$\\
                 &                              & $\sigma_2=22.4~\mev$\\
                 &                              & $f=0.12$\\
$D^-$            &  DG                          & $M_1=1864.4~\mevcc$\\
                 &                              & $\sigma_1=9.9~\mev$\\
                 &                              & $M_2=1860.6~\mevcc$\\
                 &                              & $\sigma_2=21.3~\mev$\\
                 &                              & $f=0.32$\\
 \hline \hline
\end{tabular}
\end{center}
\end{table}

\begin{figure}
  \epsfig{file=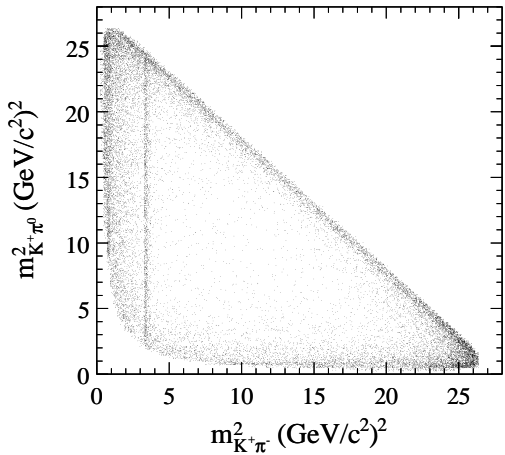,width=7cm}
  \caption {\label{fig:DalitzPlots} The DP of the selected data sample of 23,683 events. The selection criteria are described in~\secref{selection}. The $\Bz\to\overline{D}^0\piz$ decay is visible as a band near $m^2_{\Kp\pim}=3.5~(\gevcc)^2$. The remaining resonances populate the borders of the DP.}
\end{figure}

\section{THE \babar\ DETECTOR AND DATA SET}
\label{sec:DetectorAndData}

The data used in this analysis were collected with the \babar\ detector at the \pep2\ asymmetric energy \epem\ storage rings between October 1999 and September 2007. This corresponds to an integrated luminosity of 413 \invfb\ or approximately $N_{\BB}=454\pm5$ million $\BB$ pairs taken on the peak of the $\FourS$ resonance (on resonance) and 41\invfb\ recorded at a center-of-mass (CM) energy 40~$\mev$ below (off resonance).

A detailed description of the \babar\ detector is given in Ref.~\cite{NIM}. Charged-particle trajectories are measured by a five layer, double-sided silicon vertex tracker (SVT) and a 40 layer drift chamber (DCH) coaxial with a 1.5 T magnetic field. Charged particle identification is achieved by combining the information from a ring-imaging Cherenkov device (DIRC) and the ionization energy loss (\dedx) measurements from the DCH and SVT. Photons are detected, and their energies are measured in a CsI(Tl) electromagnetic calorimeter (EMC) inside the coil. Muon candidates are identified in the instrumented flux return of the solenoid.  We use GEANT4-based~\cite{GEANT} software to simulate the detector response and account for the varying beam and environmental conditions. Using this software, we generate signal and background Monte Carlo (MC) event samples in order to estimate the efficiencies and expected backgrounds in this analysis.

\section{EVENT SELECTION AND BACKGROUNDS}
\label{sec:selection}

We reconstruct $\Bz\to\Kp\pim\piz$ candidates from a $\piz$ candidate and pairs of oppositely-charged tracks that are required to form a good quality vertex. The charged-particle candidates are required to have transverse momenta above $100~\mevc$ and at least 12 hits in the DCH. We use information from the tracking system, EMC, and DIRC to select charged tracks consistent with either a kaon or pion hypothesis. The $\piz$ candidate is built from a pair of photons, each with an energy greater than $50~\mev$ in the laboratory frame and a lateral energy deposition profile in the EMC consistent with that expected for an electromagnetic shower. The invariant mass of each $\piz$ candidate, $m_{\piz}$ must be within 3 times the associated mass error, $\sigma(m_{\piz})$ of the nominal $\piz$ mass $134.9766~\mevcc$~\cite{PDG}. We also require $|\cos\theta^{*}_{\pi^0}|$, the modulus of the cosine of the angle between the decay photon and the $\piz$ momentum vector in $\piz$ rest frame, to be less than 0.95.

A $\Bz$ meson candidate is characterized kinematically by the energy-substituted mass $\mes\equiv\sqrt{(s/2+{\mathbf {p}}_i\cdot{\mathbf{p}}_B)^2/E_i^2-p_B^2}$ and the energy difference $\de \equiv E_B^*-\half\sqrt{s}$, where $(E_B,\pvec_B)$ and $(E_i,\pvec_i)$ are the four-vectors of the $B$ candidate and the initial electron-positron system in the lab frame, respectively. The asterisk denotes the \FourS\ frame, and $s$ is the square of the invariant mass of the electron-positron system. We require $5.2720~\gevcc < \mes <5.2875~\gevcc$. To avoid a bias in the DP from the dependence on the $\piz$ energy of the resolution in \de, we introduce the dimensionless quantity: 

\begin{equation}
\label{eq:deprime}
\deprime = \frac {\frac{\de} {\sigma(\de)} + m_0 + m_1x + m_2x^2 + m_3x^3}{w_0 + w_1x + w_2x^2 + w_3x^3}
\end{equation}
 
\noindent where $\sigma(\de)$ is the per event \de error and the coefficients, $m_i,w_i$ given in~\tabref{Deprimepars}, are determined from fits to signal MC in bins of $x=m_{\Kp\pim}^2~(\gevcc)^2$. We require $|\deprime| \leq 2.1$. Following the calculation of these kinematic variables, each of the $B$ candidates is refitted with its mass constrained to the world average value of the $B$-meson mass~\cite{PDG} in order to improve the DP position resolution and ensure that candidates occupy the physical region of the DP. 

\begin{table}[h]
\begin{center}
\caption{\label{tab:Deprimepars} Fitted values of $m_i$ and $w_i$ which minimize correlations of $\deprime$ with the DP position. The $m_i,w_i$ have units of $(\gevcc)^{-2i}$}
\begin{tabular}{ccc} 
\hline\hline
$i$           & $m_i~(\times 10^{-3})$    & $w_i~(\times 10^{-3})$      \\ 
\hline
0           &  $246.36$             &   $1055.09$ \\  
1           &  $-2.73$              &   $6.86$    \\
2           &  $1.14$               &   $0.66$    \\
3           &  $-0.05$              &   $-0.03$   \\
\hline\hline
\end{tabular}
\end{center}
\end{table} 

Backgrounds arise primarily from random combinations in $\ep e^-\to q\qbar$ (continuum) events. To enhance discrimination between signal and continuum, we use a neural network (NN)~\cite{NN} to combine five discriminating variables: the angles with respect to the beam axis of the $B$ momentum and $B$ thrust axis in the \FourS\ frame, the zeroth and second order monomials $L_{0,2}$ of the energy flow about the $B$ thrust axis, and $\Delta z/\sigma(\Delta z)$, the significance of the flight distance between the two \B~mesons. The monomials are defined by $ L_n = \sum_i p_i\times\left|\cos\theta_i\right|^n$, where $\theta_i$ is the angle with respect to the $B$ thrust axis of the track or neutral cluster $i$ and $p_i$ is the magnitude of its momentum. The sum excludes the tracks and neutral clusters comprising the $B$ candidate. All quantities are calculated in the \FourS\ frame. The NN is trained using off-resonance data and simulated signal events, all of which passed the selection criteria. The final sample of signal candidates is selected with a requirement on the NN output that retains $81\%$ of the signal and rejects $90\%$ of continuum events. 

Approximately 17\% of the signal MC events which have $B$ candidates passing all selection criteria except that on $\mes$, contain multiple $\B$ candidates. Since the continuum DP is modeled from the $\mes$ sideband ($5.200~\gevcc \leq \mes \leq 5.272~\gevcc$) of on-peak data, the $\mes$ requirement is not applied in selecting the best $\B$ candidate. We select the candidate with the minimum value of 

\begin{equation}
\chi^2 = \bigg({{m_{\piz} - 134.9766~\gevcc}\over{\sigma(m_{\piz})}}\bigg)^2 + \chi_{\rm Vertex}^2,  
\end{equation}

\noindent where $\chi_{\rm Vertex}^2$ is the vertex $\chi^2$ of the kinematic fit to the particles that form the \B-meson candidate.

With the above selection criteria, we determine the average signal efficiency over the DP, $\big<\epsilon\big>_{\rm DP}=21.0\pm 0.2\%$ with MC simulated data generated using the model described in Ref.~\cite{babar-kpipi}. There are 23,683 events in the data sample after the selection.

Approximately $10\%$ of the selected signal events are misreconstructed. Misreconstructed signal events, known as self-cross-feed (SCF), occur when a track or neutral cluster from the other $B$ is assigned to the reconstructed signal candidate. This occurs most often for low-momentum tracks and neutral pions; hence the misreconstructed events are concentrated in the corners of the DP. Since these regions are also where the low-mass resonances overlap significantly with each other, it is important to model the misreconstructed events correctly. We account for misreconstructed events with a resolution function described in~\secref{ML}.

MC events are used to study the background from other $B$ decays ($B$ background). We group the $B$ backgrounds into 19 different classes with similar kinematic and topological properties, collecting those decays where less than 8 events are expected into a single generic class. The $B$ background classes used in this analysis are summarized in~\tabref{bbackground}. When the yield of a class is varied in the ML fit the quoted number of events corresponds to the fit results, otherwise the expected numbers of selected events are computed by multiplying the MC selection efficiencies by the world average branching fractions~\cite{HFAG,PDG} scaled to the data set luminosity. 

The decay $\Bz\to\Kp\pim\piz$ is flavor specific (the charge of the kaon identifies the \B flavor), so the flavor of the opposite \B produced in the decay of the $\Y4S$ can be used as additional input in the analysis. Events where the opposite \B flavor has been reliably determined are less likely to be either continuum background or SCF. A neural network is trained using a large sample of MC events with ouput into 7 exclusive tagging categories identifying the flavor of the \B meson~\cite{babar-tag}. Those events where the opposite \B flavor could not be determined are included in an untagged category. Each $\Bz\to\Kp\pim\piz$ decay in the dataset is identified with the tagging category of the opposite \B determined from the neural network. 

\begin{table}[h]
\begin{center}
\caption{\label{tab:bbackground} Summary of \B background modes included in the fit model. The expected number of \B background events for each mode listed includes the branching fraction and selection efficiency. The $\Dz X$ and $\Dp X$ classes do not include $\Bz\to\overline{D}^0\piz$ and $\Bz\to\Dm\Kp$ decays which are modeled as non-interfering isobars.}

\begin{tabular}{clrr}\hline\hline
Class   & $\B$ Decay  &Events&  \\ \hline
1      & Generic                                                   &$660\pm122$& varied \\
2      & $\Dz X$                                                   &$627\pm60$ & varied \\  
3      & $\Dp X$                                                   &$370\pm80$ & varied \\ 
4      & $\Kstar(892)^0\g,\Kstar(1430)^0\g$                        &$187 \pm 14$  & \\
5      & $K^+\pim\pip$                                             &$164\pm 9$  & \\
6      & $\pip\pim\piz$                                            &$109\pm 13$  & \\
7      & $K^+\piz$                                                 &$65 \pm 3$  & \\
8      & $K^+\pim$                                                 &$53 \pm 2$  & \\  
9      & $\rho^+\rho^0$, $a_1^0\pip$, $a_1^+\piz$                   &$50\pm 13$  & \\
10     & $f_0(980)K^+$, $K^{*+}\piz$                                &$48\pm 12$  & \\
11     & $\rho^+\rho^-$, $ a_1^+\pi^-$                              &$27\pm 3$  & \\
12     & $\eta^\prime K^+$                                          &$22\pm 1$  & \\
13     & $K^{*+}\rho^-$                                             &$21\pm 6$  & \\
14     & $K^{*0}\rho^+$                                             &$15\pm 6$  & \\
15     & $K^{*+}K^-$                                                &$14\pm 10$  & \\
16     & $\rho^+\piz$                                               &$11\pm 2$  & \\
17     & $K^{*+}a_1^-$, $K^{*0}\rho^0$                               &$9\pm 2$  & \\
18     & $\pip\pim\pip$                                             &$8\pm 1$  &  \\
19     & $K^+\eta\pi^-$                                             &$8\pm 1$  & \\
\hline\hline
\end{tabular}
\end{center}
\end{table}

\section{THE MAXIMUM-LIKELIHOOD FIT}
\label{sec:ML}

We perform an unbinned extended maximum-likelihood fit to extract the $\BtoKpipi$ event yield and the resonant amplitudes. The fit uses the variables $\mes$, $\de^{\prime}$, the NN output, and the DP to discriminate signal from background. The selected on-resonance data sample consists of signal, continuum background, and \B background components. The signal likelihood consists of the sum of a correctly reconstructed (truth-matched or TM) and SCF term. The background contributions and fraction of SCF events vary with the tagging category of the opposite \B decay. We therefore separate the components of the fit by the tagging category of the opposite \B decay.

The likelihood ${\cal L}_i^\cat$ for an event $i$ in tagging category $\cat$ is the sum of the probability densities of all components,

\begin{eqnarray}
\label{eq:theLikelihood}
        {\cal L}_i^\cat
        &\equiv&
                N_{\rm sig} f^\cat_{\rm sig}
                \left[  (1-\fscfave^\cat){\cal P}_{\TM,i}^\cat +
                \fscfave^\cat{\cal P}_{\SCF,i}^\cat \right] \nonumber\\[0.3cm]
        &&
                +\; N^\cat_{q\bar q}\frac{1}{2}
                \left(1\pm\Aqq\right){\cal P}_{q\bar q,i}^\cat
               \nonumber \\[0.3cm]
        &&
                +\; \sum_{j=1}^{19}
                N_{B_j} f^\cat_{B_j}
                \frac{1}{2}\left(1\pm\ABj\right){\cal P}_{B_j,i}^\cat.
                \nonumber \\
\end{eqnarray}

\noindent Here, $j$ is the \B background class number and the $(\pm)$ is evaluated to be the charge sign of the kaon in the event $i$. A complete summary of the variables in~\equaref{theLikelihood} is given in~\tabref{DefVarLik}.

\renewcommand{\arraystretch}{1.15}
\begin{table*}[htbp]
\begin{center}
\caption{
Definitions of the different variables in the likelihood function given in~\equaref{theLikelihood}.
\label{tab:DefVarLik}}
\begin{tabular}{ll}
\hline\hline
Variable                    & Definition \\
\hline
$N_{\rm sig}$                & total number of $\Kpipi$ signal events in the data sample              \\
$f^\cat_{\rm sig}$            & fraction of signal events that are tagged in category $\cat$            \\
$\fscfave^\cat$              & fraction of SCF events in tagging category $\cat$, averaged over the DP \\
${\cal P}_{\TM,i}^\cat$       & product of PDFs of the discriminating variables used in tagging category $\cat$ for TM events  \\
${\cal P}_{\SCF,i}^\cat$      & product of PDFs of the discriminating variables used in tagging category $\cat$ for SCF events \\
$N^\cat_{q\bar q}$            & number of continuum events that are tagged in category $\cat$ \\
$\Aqq$                      & parametrizes a possible asymmetry in continuum events \\
${\cal P}_{q\bar q,i}^\cat$   & product of PDFs of the discriminating variables used in tagging category $\cat$ for continuum events \\
$N_{B_j}$                    & number of expected  $B$ background events in class $j$\\
$f^\cat_{B_j}$                & fraction of $B$ background events that are tagged in category $\cat$ \\
$\ABj$                      & parametrizes a possible asymmetry in the charged $B$ background in class $j$ \\
${\cal P}_{B_j}^\cat$        & product of PDFs of the discriminating variables used in tagging category $\cat$ for $B$ background class $j$ \\
\hline\hline
\end{tabular}
\end{center}
\end{table*}
\renewcommand{\arraystretch}{1.}

\noindent The PDFs ${\cal P}_{X}^{\cat}$~($X={\rm TM},~{\rm SCF},~q\qbar,~B_j$) are the product of the four PDFs of the discriminating variables, $d_1=\mes$, $d_2=\de^{\prime}$, $d_3 = {\rm NN~output}$, and the DP, $d_4=\{x, y\}$:

\begin{equation}
\label{eq:likVars}
        {\cal P}_{X,i}^{\cat} \;\equiv\; 
        \prod_{k=1}^4 P_{X,i}^\cat(d_k).
\end{equation}

\noindent In the fit, the DP coordinates, $(x,y)$ are transformed to square DP coordinates described in Ref.~\cite{babar-kpipi}. The extended likelihood over all tagging categories is given by

\begin{equation}
        {\cal L} \;\equiv\;  
        \prod_{\cat=1}^{7} e^{-N^\cat}\,
        \prod_{i}^{N^\cat} {\cal L}_{i}^\cat~,
\end{equation}

\noindent where $N^\cat$ is the total number of events in tagging category $\cat$. The parameterizations of the PDFs are described in~\secref{DalitzPDF} and~\secref{nondalitzPDFs}.

\subsection{The Dalitz Plot PDFs}
\label{sec:DalitzPDF}
\begin{figure}[h!]
  \begin{center}
    \includegraphics[width=7cm,keepaspectratio]{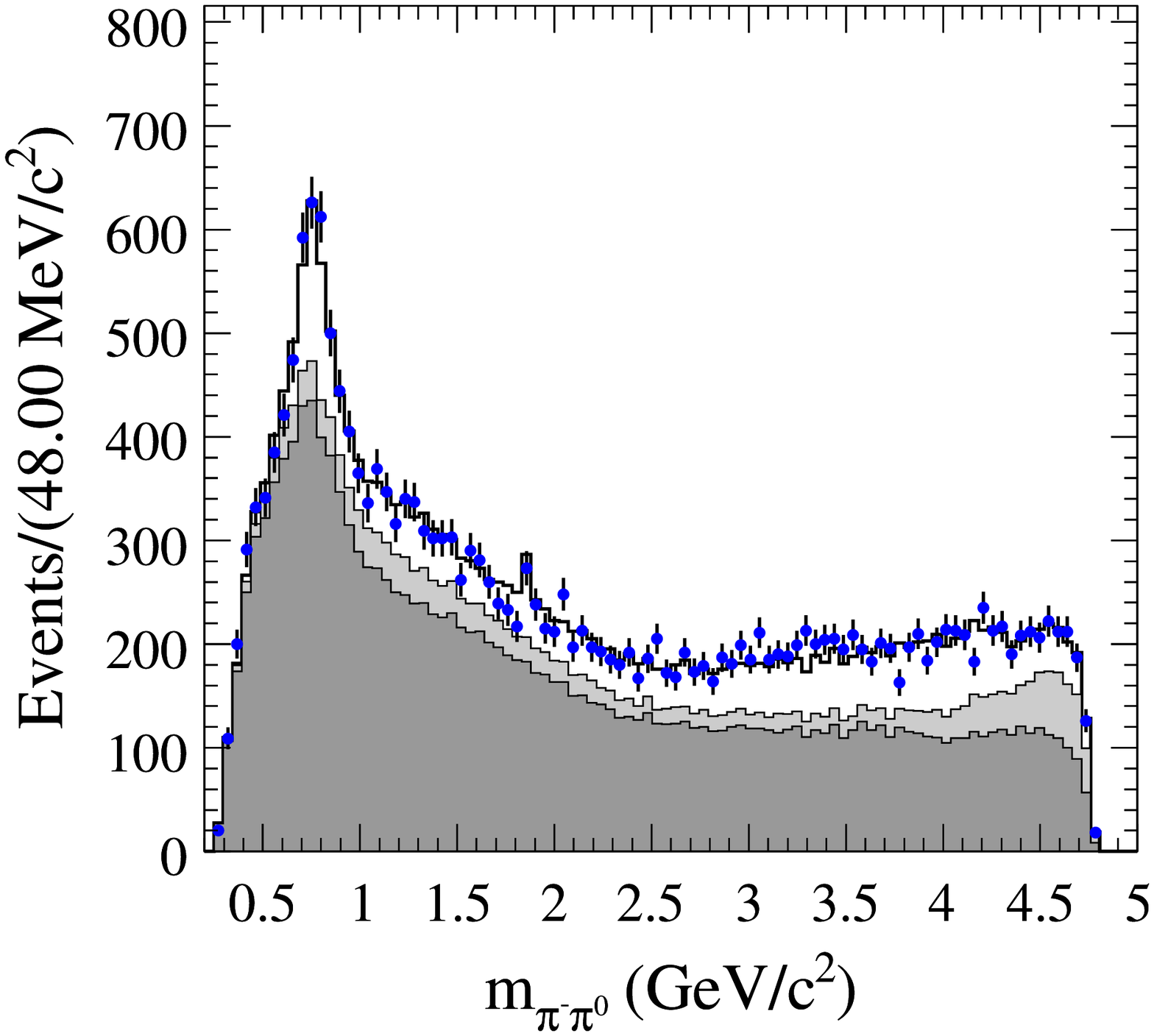} 
    \includegraphics[width=7cm,keepaspectratio]{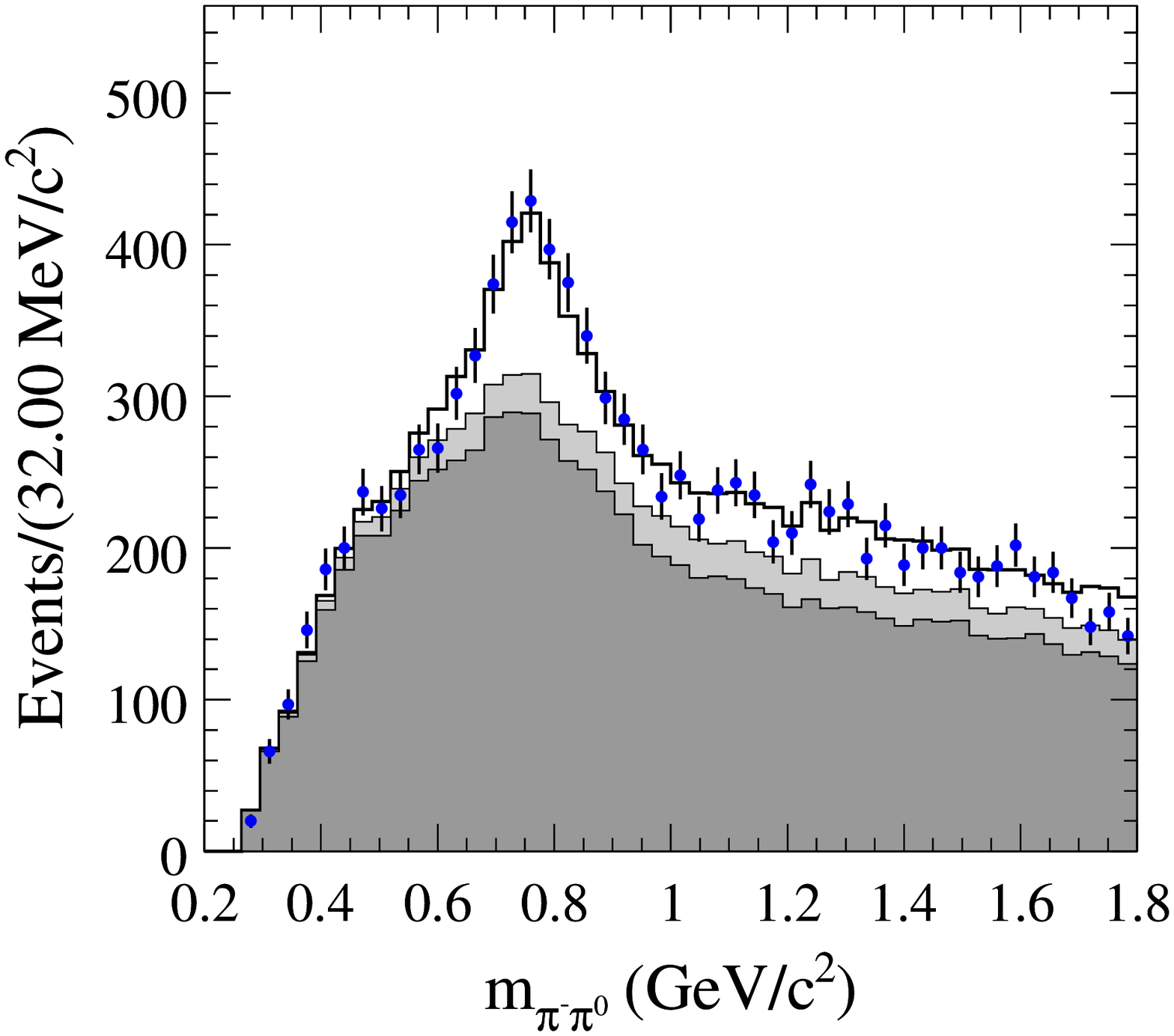} 
    \caption{\label{fig:mpimpiz}{The $m_{\pim\piz}$ invariant mass distributions in the entire kinematic range (top) and below 1.8~\gevcc (bottom) for all selected events. The $\rho(770)^-$ is visible as a broad peak near $0.77~\gevcc$. The data are shown as points with error bars. The solid histograms show the projection of the fit result for charmless signal and $D$ events (white), \B background (medium) and continuum (dark), respectively.}}
  \end{center}
\end{figure}

\begin{figure}[h!]
  \begin{center}
    \includegraphics[width=7cm,keepaspectratio]{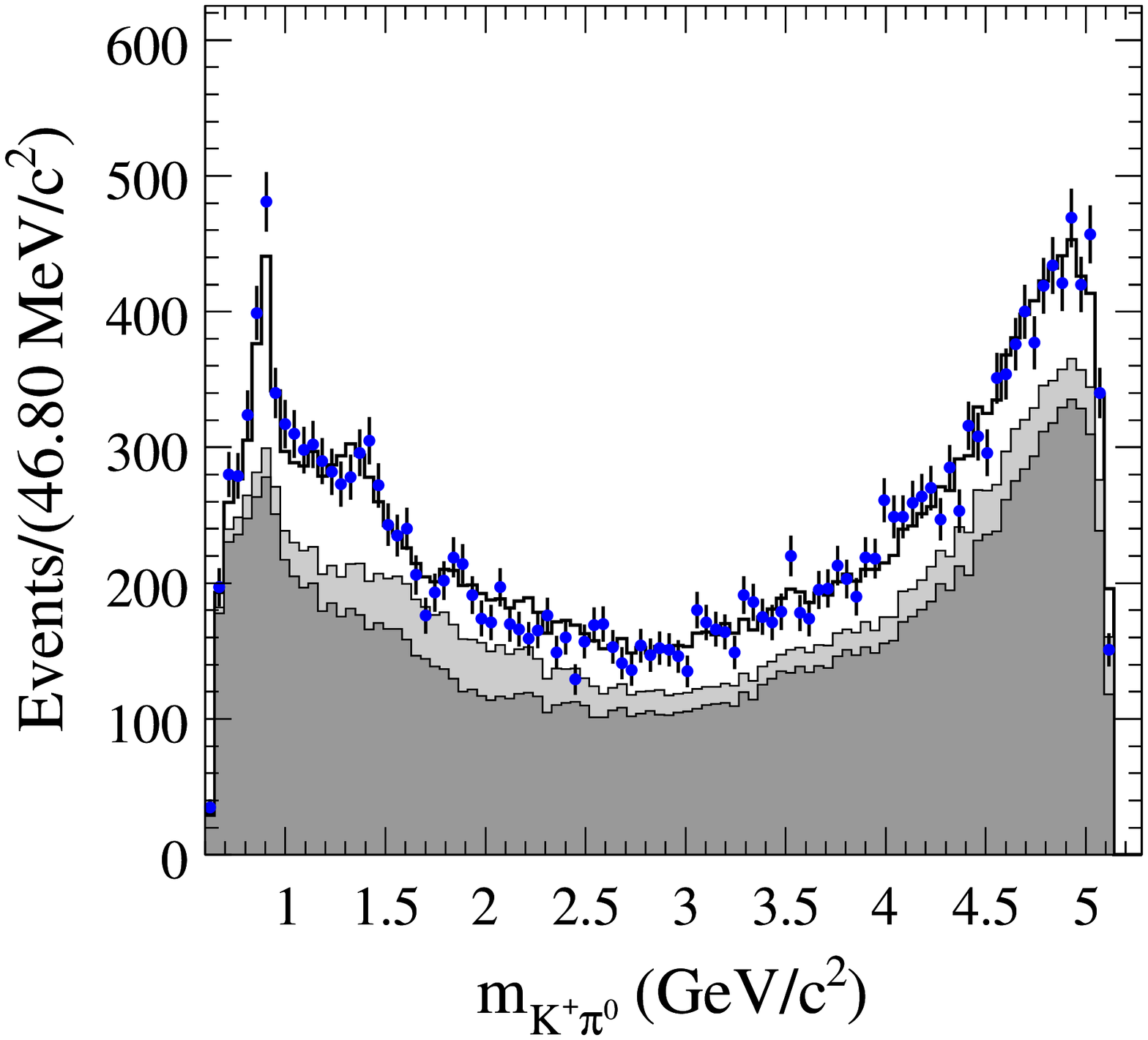} 
    \includegraphics[width=7cm,keepaspectratio]{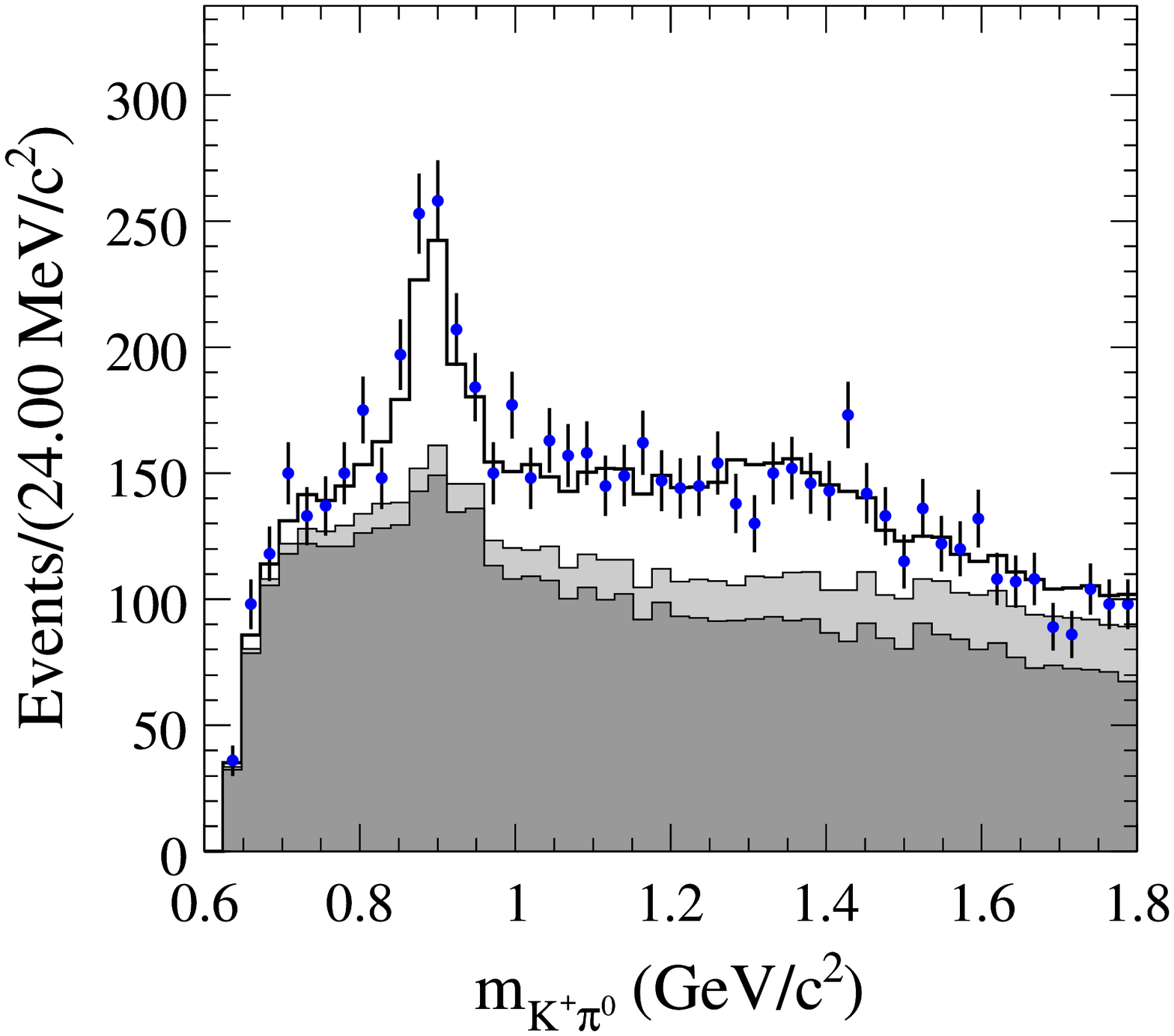}
    \caption{\label{fig:mKppiz}{The $m_{\Kp\piz}$ invariant mass distributions in the entire kinematic range (top) and below 1.8~\gevcc (bottom) for all selected events. The $\Kstar(892)^+$ is visible as a narrow peak near $0.89~\gevcc$ while the broad distribution near $1.40~\gevcc$ is the $(K\pi)_0^{*+}$. The data are shown as points with error bars. The solid histograms show the projection of the fit result for charmless signal and $D$ events (white), \B background (medium) and continuum (dark), respectively.}}
  \end{center}
\end{figure}

Since the decay $\Bz\to\Kp\pim\piz$ is flavor-specific, the \Bz and \Bzb DP distributions are independent of each other and in general can differ due to \CP~violating effects. The backgrounds, however, are largely independent of the \B flavor, hence a more reliable estimate of their contribution is obtained by fitting the \Bz and \Bzb DP distributions simultaneously. We describe only the \Bz\ DP PDF, since a change from $\mathcal{A}$ to $\mathcal{\overline{A}}$ accompanied by the interchange of the charges of the kaon and pion gives the \Bzb\ PDF. Projections of the DP are shown for each of the invariant masses $m_{\pim\piz}$ in~\figref{mpimpiz}, $m_{\Kp\piz}$ in~\figref{mKppiz}, and $m_{\Kp\pim}$ in~\figref{mKppim} along with the data. 

\subsubsection{Signal}
The total $\Bz\to\Kp\pim\piz$ amplitude is given by 

\begin{equation}
\mathcal{A}(x,y) = \sum_{k}a_ke^{i\Phi_k}f_k(J,x,y),
\end{equation}

\noindent where $k$ runs over all of components in the model described in~\secref{DecayAmplitudes}. The amplitudes and phases are measured relative to the $\rho(770)^-\Kp$ amplitude so that the phases $\Phi_{\rho(770)^-\Kp}$ and $\overline{\Phi}_{\rho(770)^+\Km}$ are fixed to 0 and the isobar, $\overline{a}_{\rho(770)^-\Kp}$ is fixed to 1. The TM signal DP PDF is 

\begin{equation}
\label{eq:PTM}
P_{\TM}(x,y) =\epsilon(x,y) (1 - \fscf(x,y))\frac{|\mathcal{A}|^2}{|N_{\TM}|^2},
\end{equation} 

\noindent where

\begin{equation}
\label{eq:NormEff}
|N_{\TM}|^2 ={\rm Re}\sum_{\kappa,\sigma}a_{\kappa}a_{\sigma}e^{i(\Phi_{\kappa} - \Phi_{\sigma})}\langle {\rm TM}| f_\kappa f^*_{\sigma}\rangle_{\rm DP},
\end{equation}

\noindent and 

\begin{equation}
  \label{eq:normAverageTM}
  \langle {\rm TM}| f_\kappa f^*_{\sigma}\rangle_{\rm DP}=\int_{\rm DP}\epsilon(x,y) (1-\fscf(x,y)) f_\kappa f^*_{\sigma}\,dx~dy.
\end{equation}

\noindent Here, $\epsilon(x,y)$ and $\fscf(x,y)$ are the DP dependent signal selection efficiency and SCF fraction. These are implemented as histograms in the square DP coordinates. The indices $\kappa$, $\sigma$ run over all components of the signal model. \equaref{normAverageTM} is evaluated numerically for the lineshapes described in~\secref{DecayAmplitudes}. 

The PDF for signal SCF is given by

\begin{equation}
\label{eq:PSCF}
P_{\SCF}(x,y) =\epsilon(x,y)\,\fscf(x,y)\frac{|\mathcal{A}|^2\otimes R_{\rm SCF}}{|N_{\SCF}|^2 \otimes R_{\rm SCF}},
\end{equation}

\noindent where $|N_{\SCF}|^2$ is given by~\equaref{NormEff} with the replacement ${\rm TM}\to {\rm SCF}$, and 

\begin{equation}
  \label{eq:normAverageSCF}
  \langle {\rm SCF}| f_\kappa f^*_{\sigma}\rangle_{\rm DP}=\int_{\rm DP}\epsilon(x,y)\fscf(x,y) f_\kappa f^*_{\sigma}\,dx~dy.
\end{equation}

\noindent Convolution with a resolution function is denoted by $\otimes R_{\rm SCF}$. In contrast with TM events, a convolution is necessary for SCF, since mis-reconstructed events often incur large migrations over the DP, \ie~the reconstructed coordinates $(x_r,y_r)$ are far from the true values $(x_t,y_t)$. This can correspond to a broadening of resonances by as much as $800~\mev$. We introduce a resolution function, $R_{\rm SCF}(x_r,y_r;x_t,y_t)$, which represents the probability to reconstruct at the coordinates $(x_r,y_r)$ a SCF event that has the true coordinate $(x_t,y_t)$. The resolution function is normalized so that

\begin{equation}
\label{eq:theMatrix}
\int_{\rm DP}R_{\rm SCF}(x_r,y_r;x_t,y_t)~dx_r~dy_r =1~\forall~(x_t,y_t), 
\end{equation}

\noindent and is implemented as an array of 2-dimensional histograms that store the probabilities as a function of the DP coordinates. $R_{\rm SCF}$ is convolved with the signal model in the expression of $\mathcal{P}_{\rm SCF}$ in~\equaref{PSCF} to correct for broadening of the SCF events.

The normalization of the total signal PDF is guaranteed by the DP-averaged fraction of SCF events, 

\begin{equation}
\label{eq:fscfAverage}
\fscfave^\cat \;=\; \frac{\int_{\rm DP} \epsilon(x,y)\,\fscf^\cat(x,y) (\AmpAll)\,dx~dy}{\int_{\rm DP} \epsilon(x,y) (\AmpAll)\,dx~dy}~.
\end{equation}

\noindent This quantitiy is decay dynamics-dependent, and in principle must be computed iteratively. Typically, $\fscfave^\cat\approx9\%$ converging rapidly after a small number of fits.

\subsubsection{Background}
The continuum DP distribution is extracted from a combination of off-resonance data and an $\mes$ sideband ($5.200~\gevcc \leq \mes \leq 5.272~\gevcc$) of the on-resonance data from which the \B background has been subtracted. The DP is divided into eight regions where different smoothing parameters are applied in order to optimally reproduce the observed wide and narrow structures by using a two-dimensional kernel estimation technique~\cite{Cranmer}. A finely binned histogram is used to describe the peak from the narrow $\Dz$ continuum production. Most \B background DP PDFs are smoothed two-dimensional histograms obtained from MC. The backgrounds due to $b\to c$ decays with $\Dz$ mesons (class $2$ in~\tabref{bbackground}), are modeled with a finely binned histogram around the $\Dz$ mass.

\begin{figure}[h!]
  \begin{center}
    \includegraphics[width=7cm,keepaspectratio]{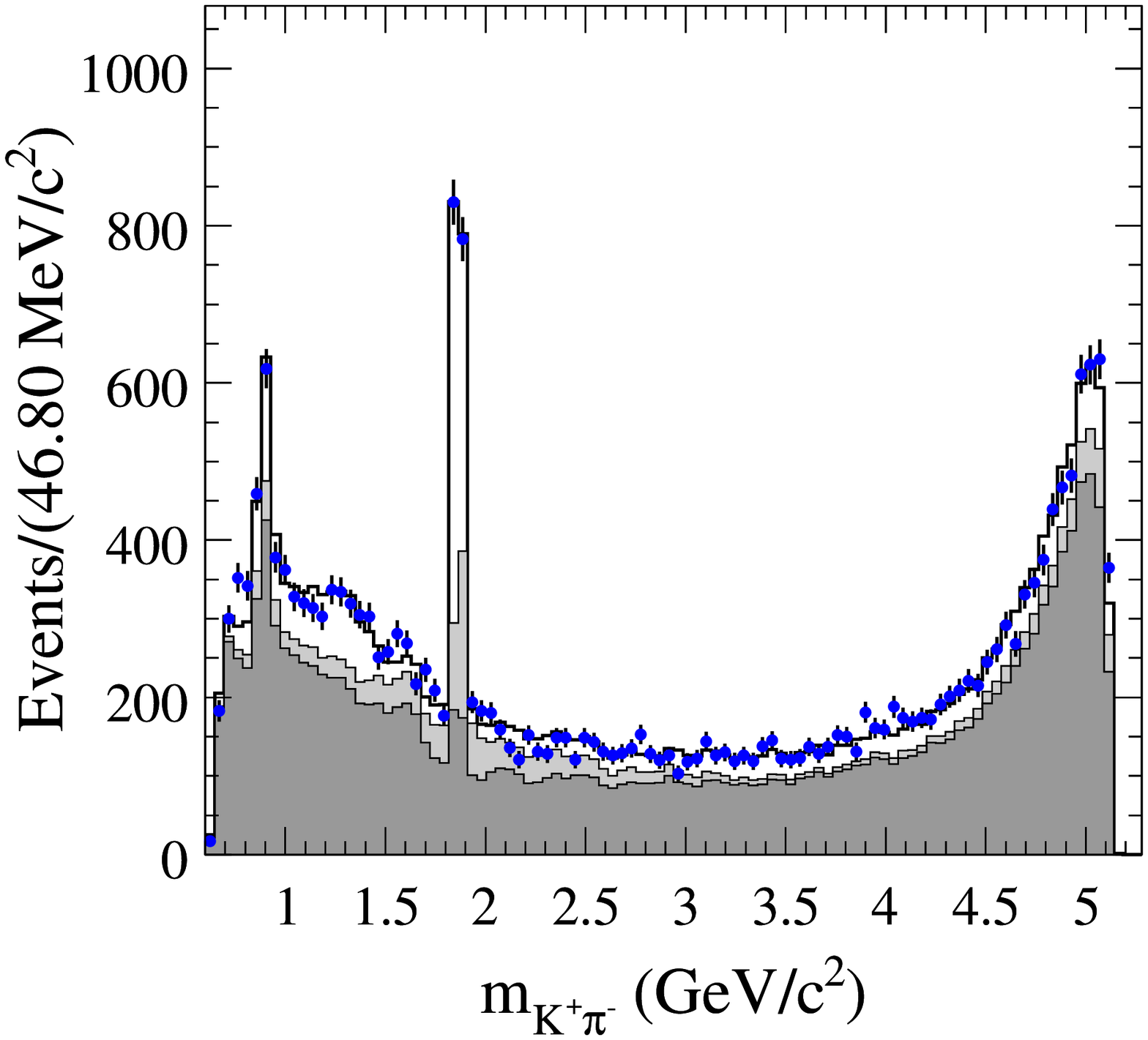} 
    \includegraphics[width=7cm,keepaspectratio]{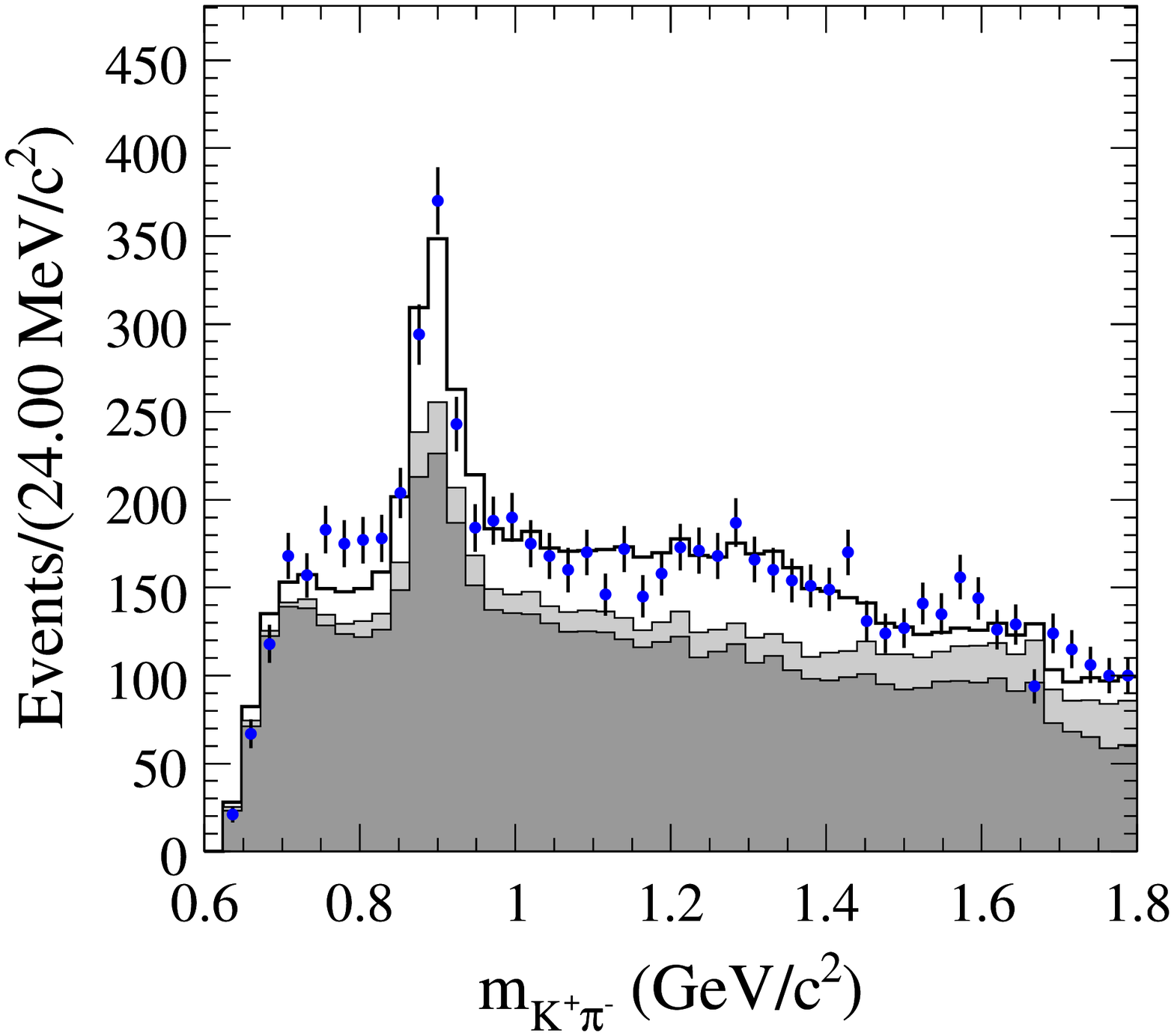}
    \caption{\label{fig:mKppim}{The $m_{\Kp\pim}$ invariant mass distributions in the entire kinematic range (top) and below 1.8~\gevcc (bottom) for all selected events. The $\Kstar(892)^0$ is visible as a narrow peak near $0.89~\gevcc$ while the broad distribution near $1.40~\gevcc$ is the $(K\pi)_0^{*0}$. The narrow peak near $1.8~\gevcc$ is the $\Dz$ meson. The data are shown as points with error bars. The solid histograms show the projection of the fit result for charmless signal and $D$ events (white), \B background (medium) and continuum (dark), respectively.}}
  \end{center}
\end{figure}

\subsection{Description of the Other Variables}
\label{sec:nondalitzPDFs}
\subsubsection{Signal}

The $\mes$ distribution for TM-signal events is parameterized by a modified Gaussian distribution given by

\begin{equation}
  \label{eq:cruijff}
  P_{\rm sig-TM}(d_1)=\exp\left[ -\frac{(d_1-m)^2} {2\sigma^2_\pm+\alpha_\pm(d_1-m)^2}\right].
\end{equation}

\noindent The peak of the distribution is given by $m$ and the asymmetric width of the distribution $\sigma_\pm$ is given by $\sigma_+$ for $d_1 < m$ and $\sigma_-$ for $m < d_1$. The asymmetric modulation $\alpha_\pm$ is similarly given by $\alpha_+$ for $d_1 < m$ and $\alpha_-$ for $m < d_1$. The parameters in~\equaref{cruijff} are determined in the data fit. The $\mes$ distribution for SCF-signal events is a smoothed histogram produced with a Gaussian kernel estimation technique from MC. 

The \deprime\ distribution for TM-signal is parameterized by the sum of a Gaussian and a first order polynomial PDF,

\begin{equation}
  \label{eq:GP}
  P_{\rm sig-TM}(d_2)=\frac{1}{\sigma}\exp\left[-\frac{(d_2-m)^2}{2\sigma^2}\right] + ad_2 + b.
\end{equation}

\noindent The parameters $m,~\sigma,~a$ given in~\equaref{GP} are described by linear functions of $x=m_{\Kpm\pimp}^2$ with slopes and intercepts determined in the fit in order to account for a small residual correlation of \deprime\ with the DP position. A smoothed histogram taken from MC is used for the SCF-signal \deprime\ PDF. The NN PDFs for signal events are smoothed histograms taken from MC.
 
\subsubsection{Background}
We use the ARGUS function~\cite{Argus}

\begin{equation}
\label{eq:argus}
P_{q\qbar}\left(z={\frac{d_1}{\mes^{\rm max}}}\right)\propto z\sqrt{1-z^2} e^{-\xi(1-z^2)},
\end{equation}

\noindent as the continuum \mes\ PDF. The endpoint $\mes^{\rm max}$ is fixed to $5.2897~\gevcc$ and $\xi$ is determined in the fit. The continuum \deprime\ PDF is a linear function with slope determined in the fit. The shape of the continuum NN distribution is correlated with the DP position and is described by a function that varies with the closest distance between the point representing the event and the boundary of the DP $\Delta_{\mathrm{DP}}$,

\begin{equation}
    \label{eq:NNDP}
    P_{q\qbar}(d_3;\Delta_{\mathrm{DP}}) = (1-d_3)^{k_1} (k_2 d_3^2 + k_3 d_3 + k_4).
\end{equation}

\begin{figure*}[htb]
  \includegraphics[width=5.5cm,keepaspectratio]{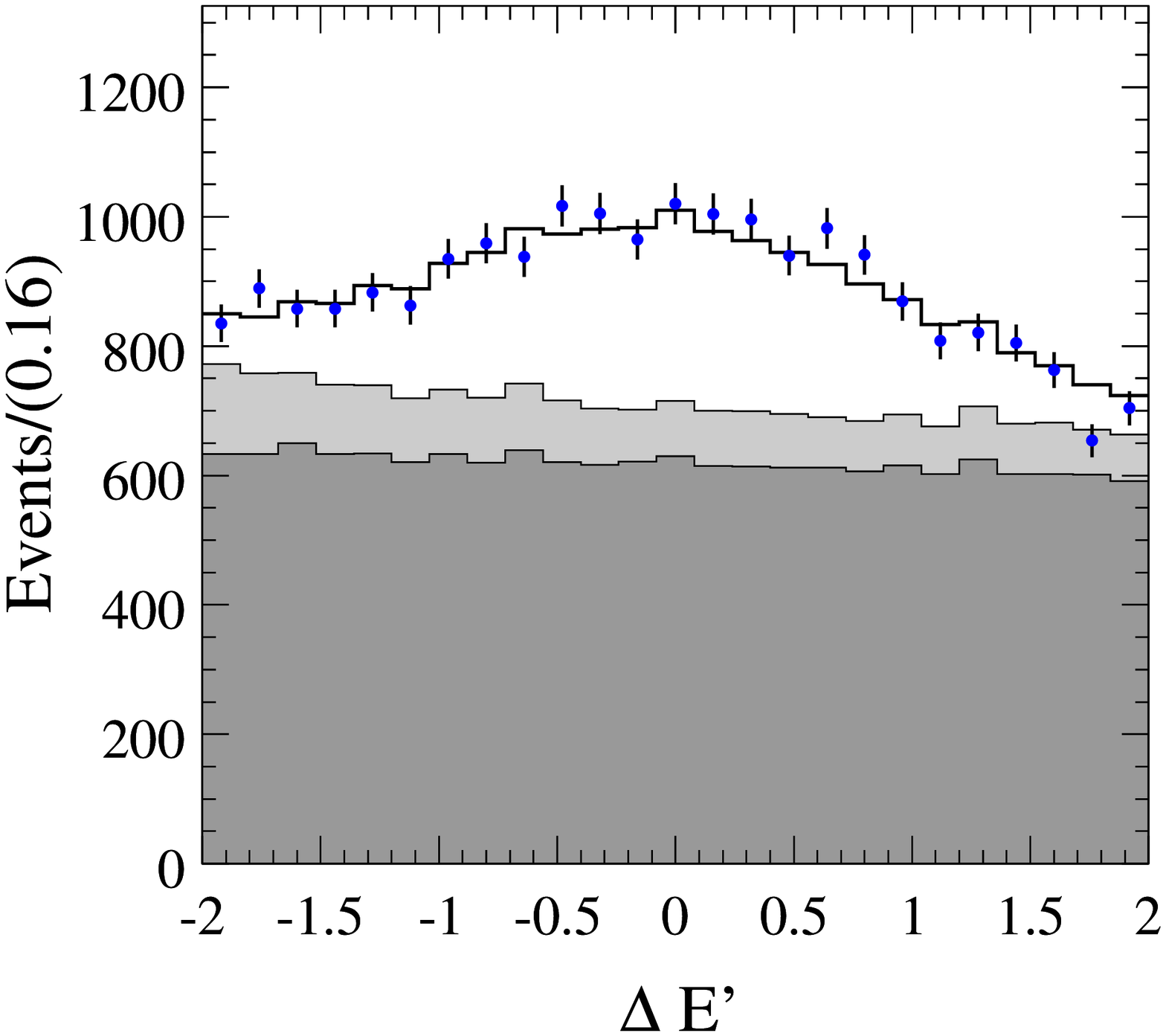}
  \includegraphics[width=5.5cm,keepaspectratio]{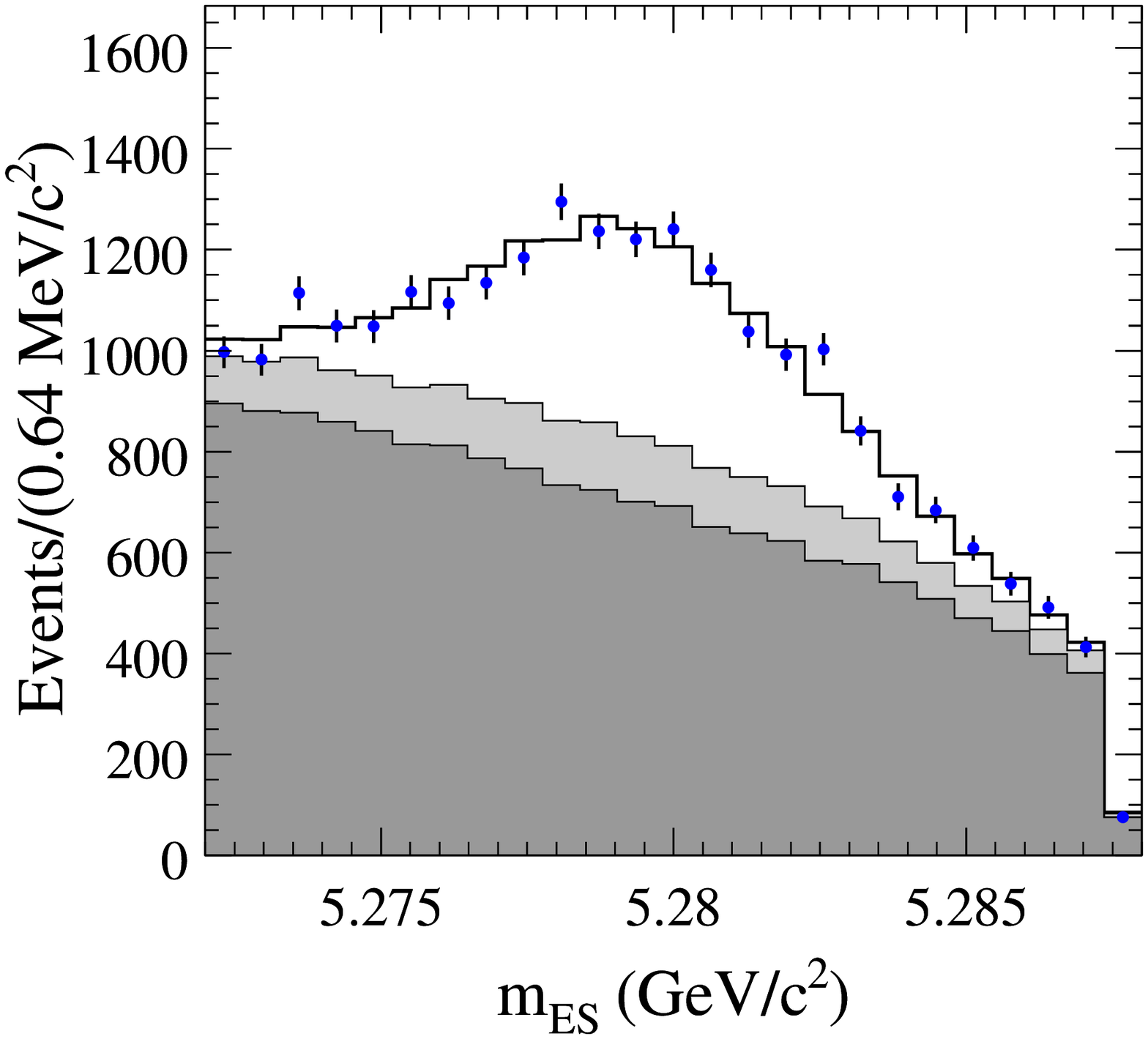}
  \includegraphics[width=5.5cm,keepaspectratio]{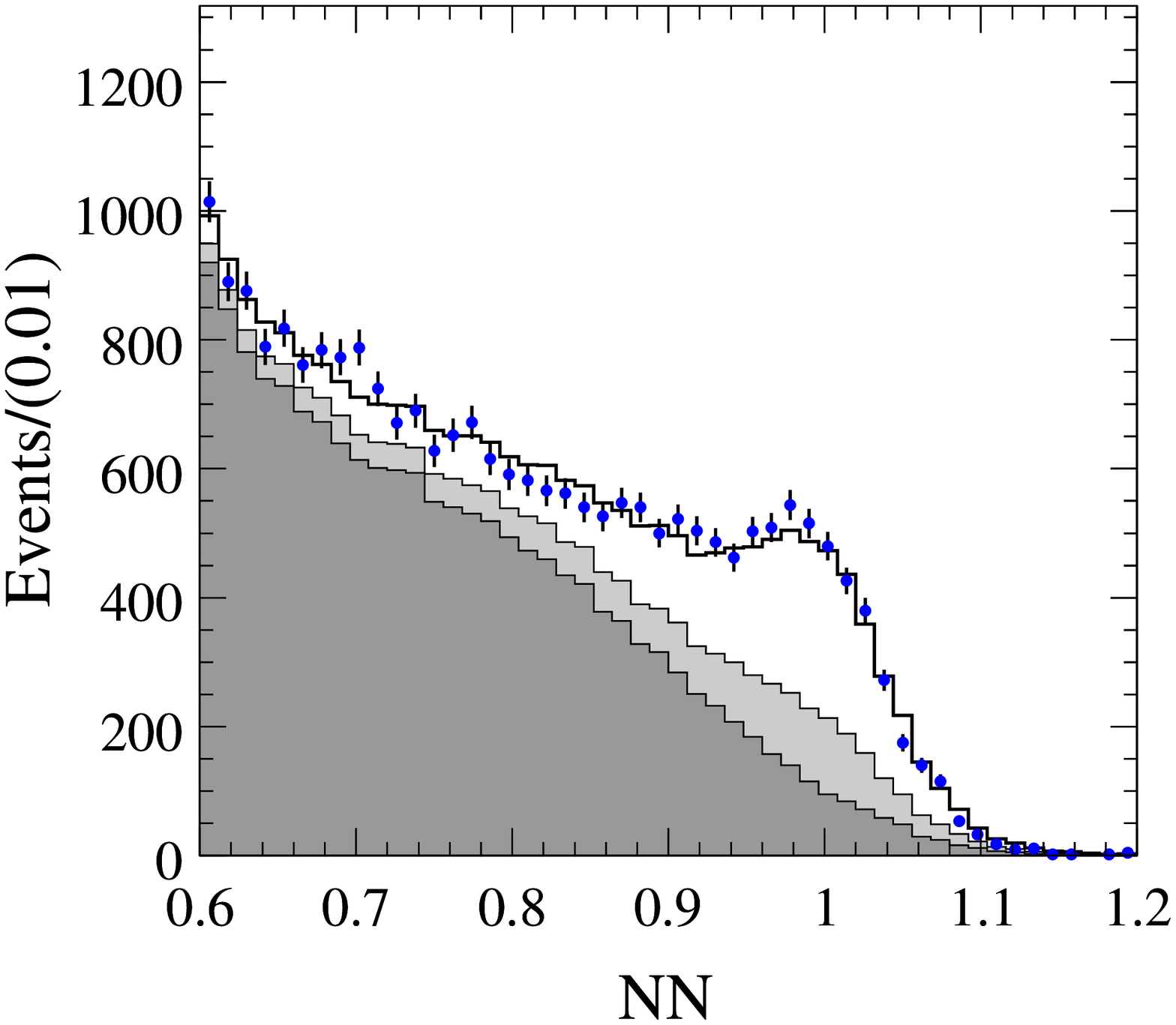}
  \caption{\label{fig:projections} Distributions of $\de^{\prime}$ (left), $\mes$ (center), and $NN$ output (right) for all selected events. The data are shown as points with error bars. The solid histograms show the projection of the fit result for charmless signal and $D$ events (white), \B background (medium) and continuum (dark), respectively.}
\end{figure*}

\noindent Here, $k_i = q_i + p_i\cdot \Delta_{\mathrm{DP}}$ where $q_i,p_i$ are determined in the fit. We use smoothed histograms taken from MC to describe \mes, \deprime~and NN distributions for the \B background classes in~\tabref{bbackground}. Projections of the \mes, \deprime~and NN PDFs are shown in~\figref{projections} for signal, \B background and continuum events along with the data.

\section{RESULTS}
\label{sec:Results}

The ML fit results in a charmless signal yield of $N_{\rm sig} = 3670\pm96~(\rm{stat.})\pm94~(\rm{syst.})$ events and total branching fraction for charmless $\Bz\to\Kp\pim\piz$ decays of ${\cal B}_{\rm sig} = 38.5\pm 1.0~({\rm stat.}) \pm 3.9~({\rm syst.})\times 10^{-6}$. We find the yields for $\Bz\to\overline{D}^0\piz$ and $\Bz\to D^-\Kp$ events are consistent with the expectations based on their world average branching fractions. The sources of systematic uncertainty, including those related to the composition of the DP, are discussed in~\secref{Systematics}. When the fit is repeated starting from input parameter values randomly distributed within the statistical uncertainty of the values obtained from the ML fit for the magnitudes, and within the [$-\pi, \pi$] interval for the phases, we observe convergence toward four minima of the negative log-likelihood function (${\rm NLL}=-{\rm log}~{\cal L}$). The best solution is separated by 5.4 units of NLL ($3.3~\sigma$) from the next best solution. The event yield we quote is for the best solution; the spread of signal yields between the four solutions is less than 5 events. The phases $\Phi$ and $\overline{\Phi}$, \CP~asymmetries and branching fractions determined by the ML fit are given for the best solution in~\tabref{bestsolution}. We quote the total branching fractions in~\tabref{bestsolution} assuming all $\Kstar\to K\pi$ and $\rho\to\pip\pim$ branching fractions to be 100\% and isospin conservation in $\Kstar\to K\pi$ decays. In the Appendix we list the fitted magnitudes and phases for the four solutions in~\tabref{fourfitsolutions} and together with the correlation matrix for the best solution in Tables~\ref{tab:Corr_Amp_I}-\ref{tab:Corr_Ampbar_II}. 

\begin{table*}[htb]
\begin{center}
\caption{\label{tab:bestsolution}
\CP-averaged branching fractions ${\cal B}_k$, phases $\overline{\Phi}$ and $\Phi$ for $\Bzb$ and $\Bz$ decays respectively, measured relative to $\Bz(\Bzb)\to\rho(770)^\mp K^\pm$, and \CP~asymmetries, $A_{C\!P}$ defined in~\equaref{PartialFractions}. The first error is statistical and the second is systematic. When the elastic range term is separated from the $K\pi$ S-wave we determine the total NR branching fraction ${\cal B}_{\rm NR} = 7.6\pm 0.5~({\rm stat.})\pm 1.0~({\rm syst.}) \times 10^{-6}$ and the resonant $\Kstar_0(1430)\pi$ branching fractions ${\cal B}_{\Kstar_0(1430)^+\pim} = 27.8\pm 2.5~({\rm stat.})\pm 3.3~({\rm syst.}) \times 10^{-6}$, ${\cal B}_{\Kstar_0(1430)^0\piz} = 7.0\pm 0.5~({\rm stat.})\pm 1.1~({\rm syst.}) \times 10^{-6}$. }
\setlength{\tabcolsep}{.9pc}
\begin{tabular}{ccccc}
\hline \hline\\
 Isobar       & ${\cal B}$ ($\times 10^{-6}$)            &  $\overline{\Phi}~[^{\circ}]$   &  $\Phi~[^{\circ}]$             & $A_{C\!P}$                      \\ 
\hline \\
$\rho(770)^{-}\Kp$ 	& $ 6.6 \pm 0.5 \pm 0.8    $   & $ 0~(\mathrm{fixed})$       &  $ 0~(\mathrm{fixed})  $         & $ 0.20  \pm 0.09  \pm 0.08 $      \\
$\rho(1450)^{-}\Kp$ 	& $ 2.4 \pm 1.0 \pm 0.6    $   & $ 75 \pm 19 \pm 9   $&  $ 126  \pm 25  \pm 26 $                & $ -0.10 \pm 0.32  \pm 0.09 $      \\
$\rho(1700)^{-}\Kp$ 	& $ 0.6 \pm 0.6 \pm 0.4    $   & $ 18 \pm 36 \pm 16  $&  $ 50   \pm 38  \pm 20 $                & $ -0.36 \pm 0.57  \pm 0.23 $      \\
$\Kstar(892)^+\pim$ 	& $ 8.0 \pm 1.1 \pm 0.8    $   & $ 33 \pm 22 \pm 20  $&  $ 39   \pm 25  \pm 20 $                & $ -0.29 \pm 0.11  \pm 0.02 $      \\
$\Kstar(892)^0\piz$ 	& $ 3.3 \pm 0.5 \pm 0.4    $   & $ 29 \pm 18 \pm 6   $&  $ 17   \pm 20  \pm 8  $                & $ -0.15 \pm 0.12  \pm 0.04 $      \\
$(K\pi)^{*+}_0\pim$ 	& $ 34.2\pm 2.4 \pm 4.1    $   & $-167\pm 16 \pm 37  $&  $ -130 \pm 22  \pm 22 $                & $  0.07 \pm 0.14  \pm 0.01 $      \\
$(K\pi)^{*0}_0\piz$ 	& $ 8.6 \pm 1.1 \pm 1.3    $   & $ 13 \pm 17 \pm 12  $&  $ 10   \pm 17  \pm 16 $                & $ -0.15 \pm 0.10  \pm 0.04 $      \\
NR 			& $ 2.8 \pm 0.5 \pm 0.4    $   & $ 48 \pm 14 \pm 6   $&  $ 90   \pm 21  \pm 15 $                & $  0.10 \pm 0.16  \pm 0.08 $      \\
\hline \hline\\
\end{tabular}
\end{center}
\end{table*}

We measure the relative phase between the narrow $\Bz\to\Kstar(892)^+\pim$ and $\Bz\to\Kstar(892)^0\piz$ resonances despite their lack of kinematic overlap due to significant contributions from the $\Bz\to(K\pi)_0^{*0}\piz,~(K\pi)_0^{*+}\pim$ S-waves, $\Bz\to\rho(770)^-\Kp$ and NR components. Each of these components interferes with both the $\Bz\to\Kstar(892)^+\pim$ and $\Bz\to\Kstar(892)^0\piz$ resonances, providing a mechanism for their coherence. The relative phases among the resonances are consequently sensitive to the models for their kinematic shapes. We discuss the systematic uncertainty associated with mismodeling of the resonance shapes in~\secref{Systematics}.   

The quality of the fit to the DP is appraised from a $\chi^2$ of 745 for 628 bins where at least 16 events exist in each bin. The relatively poor fit appears to be due to mismodeling of the continuum background which comprises $69\%$ of the 23,683 events. A signal enhanced subsample of 3232 events is selected by requiring the signal likelihood of events to be greater than $20\%$ as determined by the product of the NN, $\deprime$, and $\mes$ PDFs. Using the signal enhanced subsample we find a $\chi^2$ of 149 for 140 bins where at least 16 events exist in each bin. The excess events near $0.8~\gevcc$ seen in~\figref{mKppim} are not observed in the signal enhanced subsample. The systematic uncertainty associated with continuum mismodeling is described in~\secref{Systematics}. 

\begin{figure}[h!]
  \begin{center}
    \includegraphics[width=7cm,keepaspectratio]{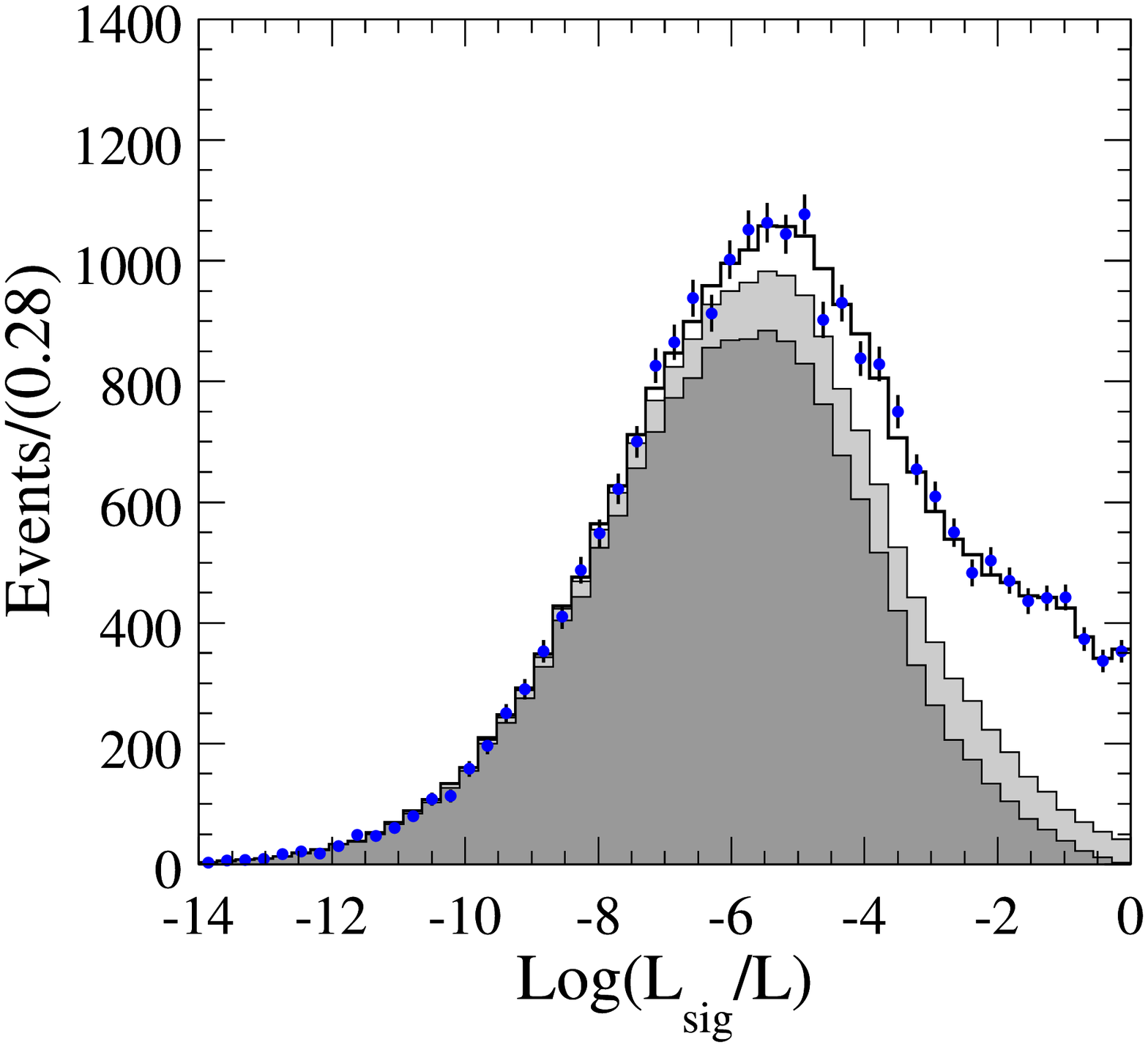} 
    \includegraphics[width=7cm,keepaspectratio]{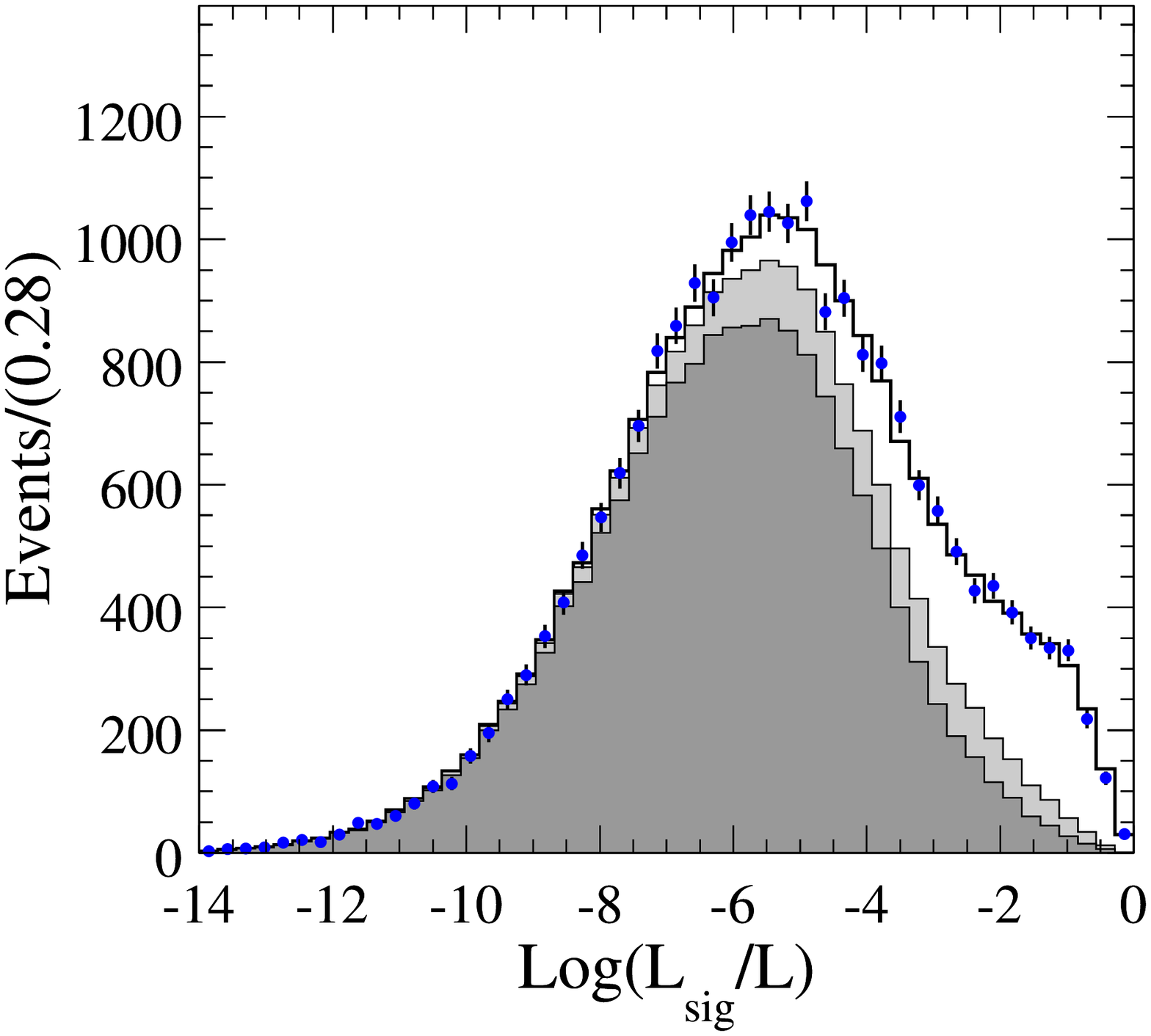}
    \caption{\label{fig:Likelihoodplots} Distributions of the log-likelihood for all events (top) and for events excluding the $D^0$ region $1.8~\gevcc < m_{\Kp\pim} < 1.9~\gevcc$ (bottom). The data are shown as points with error bars. The solid histograms show the projection of the fit result for charmless signal and $D$ events (white), \B background (medium) and continuum (dark), respectively.}
  \end{center}
\end{figure}

We validate the fit model by generating 100 data sample sized pseudo-experiments with the same isobar values as the best solution, and observe that the NLL of the data fit falls within the NLL distribution of the pseudo-experiments. The distributions of log-likelihood ratio, ${\rm log}({\cal L_{\rm sig}}/{\cal L})$~(see~\equaref{theLikelihood}) are shown in~\figref{Likelihoodplots}. The distributions show good agreement of the data with the fit model. The agreement remains good when events near the $D^0$ region of the DP ($1.8~\gevcc < m_{\Kp\pim} < 1.9~\gevcc$) are removed from the log-likelihood distribution.  

\section{SYSTEMATIC UNCERTAINTIES}
\label{sec:Systematics}

Since the amount of time required for the likelihood fit to converge dramatically increases with the number of isobar parameters to determine, we limit our isobar model to only those resonances with significant branching fractions. The dominant systematic uncertainty in this analysis is due to contributions from intermediate resonances not included in the isobar model. We include the $\Kstar(1680)\pi$ and tensor $\Kstar_2(1430)\pi$ resonances with line shapes described in~\tabref{extrares} in a fit to data. The result of this fit is used to generate high statistics samples of MC including these resonances. These samples are then fitted with the nominal isobar model and the observed shifts in the isobar parameters are taken as a systematic uncertainty listed in the Isobar Model field in Tables~\ref{tab:systTable_K}, and~\ref{tab:systTable_rho}. We find the $\Kstar(1680)\pi$ and $\Kstar_2(1430)\pi$ amplitudes each to contribute less than $1\%$ of the signal yield. 

\begin{table}[h]
\begin{center}
\caption{\label{tab:extrares} The line shape parameters of the additional $\Kstar_2(1430)^0,\ \Kstar_2(1430)^+$, $\Kstar(1680)^0$ and $\Kstar(1680)^+$ resonances.}
\begin{tabular}{ccl} \hline\hline
Resonance & Lineshape & Parameters  \\ \hline
\multicolumn{3}{c}{{Spin $J=1$}}\\ 
&& \\
$\Kstar(1680)^0$           & RBW~\cite{PDG} & $M=1717~\mevcc$  \\                                                           
                           &     & $\Gamma^0=322~\mev$\\ 
                           &     & $R=1.5~(\gev)^{-1}$ \\
$\Kstar(1680)^+$           & RBW~\cite{PDG} & $M=1717~\mevcc$ \\
                           &     & $\Gamma^0=322~\mev$ \\
                           &     & $R=1.5~(\gev)^{-1}$ \\
\multicolumn{3}{c}{{Spin $J=2$}}\\ 
&& \\
$\Kstar_2(1430)^0$         & RBW~\cite{PDG}  & $M=1432~\mevcc$ \\
                           &      & $\Gamma^0=109~\mev$ \\
                           &      & $R=1.5~(\gev)^{-1}$ \\
$\Kstar_2(1430)^+$         & RBW~\cite{PDG}  & $M=1425~\mevcc$ \\
                           &      & $\Gamma^0=98.5~\mev$ \\
                           &      & $R=1.5~(\gev)^{-1}$ \\
 \hline \hline
\end{tabular}
\end{center}
\end{table}

Mis-modeling of the continuum DP (CDP) distribution is the second most significant source of systematic uncertainty in the isobar parameters of the signal DP model. Due to the limited amount of off-resonance events recorded at $\babar$ the CDP distribution is modeled from the \mes sideband as described in~\secref{selection}. Events in the \mes sideband have necessarily higher momentum than those near the signal peak and hence have a different DP distribution. To quantify the effect of modeling the \mes on-resonance CDP with off-resonance events we use a high statistics sample of $\qqbar$ MC to create a model of the CDP from the \mes signal region. We then generate 100 pseudo-experiments with the \mes signal region CDP model and fit each of these with both the on-resonance and off-resonance models of the CDP. The average difference observed in the isobar parameters between fits with each of the CDP models is recorded in the Continuum DP field of Tables~\ref{tab:systTable_K} and~\ref{tab:systTable_rho}. In order to quantify the effect of mis-modeling of the shape of the continuum DP with the nominal smoothing parameter, we recreate the continuum DP PDF with various smoothing parameters. We refit the data using these alternate continuum DP PDFs and record the variations in the isobars in the PDF shape parameter field of Tables~\ref{tab:systTable_K} and~\ref{tab:systTable_rho}. 

Other sources of systematic uncertainty include: the uncertainty in the masses and widths of the resonances, the uncertainty in the fixed \B background yields, the mis-estimation of SCF and identification efficiencies in MC, and a small intrinsic bias in the fitting procedure. We vary the masses, widths and other resonance parameters within the uncertainties quoted in~\tabref{nominal}, and assign the observed differences in the measured amplitudes as systematic uncertainties (Lineshape field in Tables~\ref{tab:systTable_K} and~\ref{tab:systTable_rho}). To estimate the systematic uncertainty due to fixing the \B background yields, we float each of the \B background contributions in a series of fits to data. We record the variation in the isobar parameters in the \B background field of Tables~\ref{tab:systTable_K} and~\ref{tab:systTable_rho}. 

The average fraction of misreconstructed signal events predicted by MC has been verified with fully reconstructed $\B\to D\rho$ events~\cite{babar-scf}. No significant differences between data and the simulation are found. We estimate the effect of misestimating the SCF fractions in MC by varying their nominal values relatively by $\pm10\%$ in a pair of fits to data. The average shift in the isobar parameters is recorded in the SCF fraction field of Tables~\ref{tab:systTable_K} and~\ref{tab:systTable_rho}. 

\section{INTERPRETATION}
\label{sec:Interpretation}
Here, we use the results of this analysis and that presented for $\Bz\to \KS\pip\pim$~\cite{babar-kspipi} to construct isospin amplitudes as described by~\equaref{A32} and~\equaref{A32_rhoK}. Individually, the phases of these amplitudes provide sensitivity to the CKM angle $\gamma$ while together they have been shown to obey the sum rule defined in~\equaref{sumrule} in the limit of SU(3) symmetry. 

\subsection{Measurement of $\Phi_{3\over2}$}
\label{sec:Phi32}
\subsubsection{$\B\to\Kstar\pi$ decays}

The weak phase of ${\cal A}_{3\over2}$ in~\equaref{A32}, expressed as a function of the phases and magnitudes of isobar amplitudes is given by~\cite{Wagner}, 

\begin{equation}
\label{eq:Phi3half}
\Phi_{3\over2} = {1\over2} (\delta - \overline{\delta} + \Delta{\phi_{K^*\pi}}).
\end{equation}  

\noindent Here,

\begin{equation}
\label{eq:phi3half}
\deltaordeltabar = \arctan \Bigg({{\sqrt{2}|\AorAbar_{\Kstarz\piz}|\sorsbar}\over{|\AorAbar_{\Kstarp\pim}| + \sqrt{2}|\AorAbar_{\Kstarz\piz}|\corcbar}} \Bigg),
\end{equation}

\noindent and $\AorAbar_{\Kstarz\piz}$, $\AorAbar_{\Kstarp\pim}$ are the isobar amplitudes given in~\equaref{isobars}. We define 

\begin{eqnarray}
\sorsbar &=& \sin{(\PhiorPhibar_{\Kstarz\piz} - \PhiorPhibar_{\Kstarp\pim})}, \\
\corcbar &=& \cos{(\PhiorPhibar_{\Kstarz\piz} - \PhiorPhibar_{\Kstarp\pim})}, \\
\Delta{\phi_{\Kstar\pi}} &=& \Phi_{\Kstarp\pim} - \overline{\Phi}_{\Kstarm\pip}. 
\end{eqnarray}

\begin{figure}[h!]
  \begin{center}
    \epsfig{file=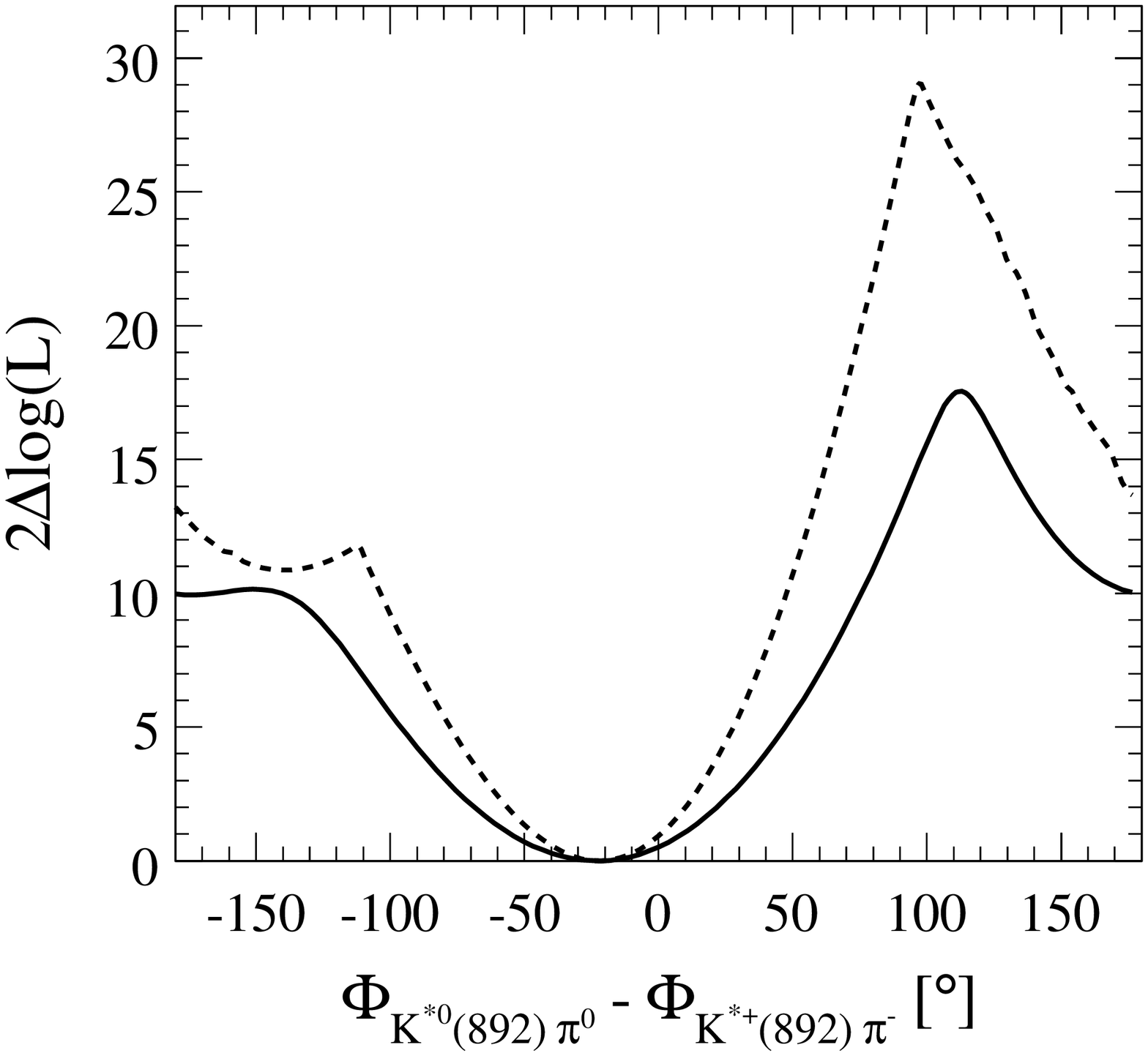,width=7cm} 
    \epsfig{file=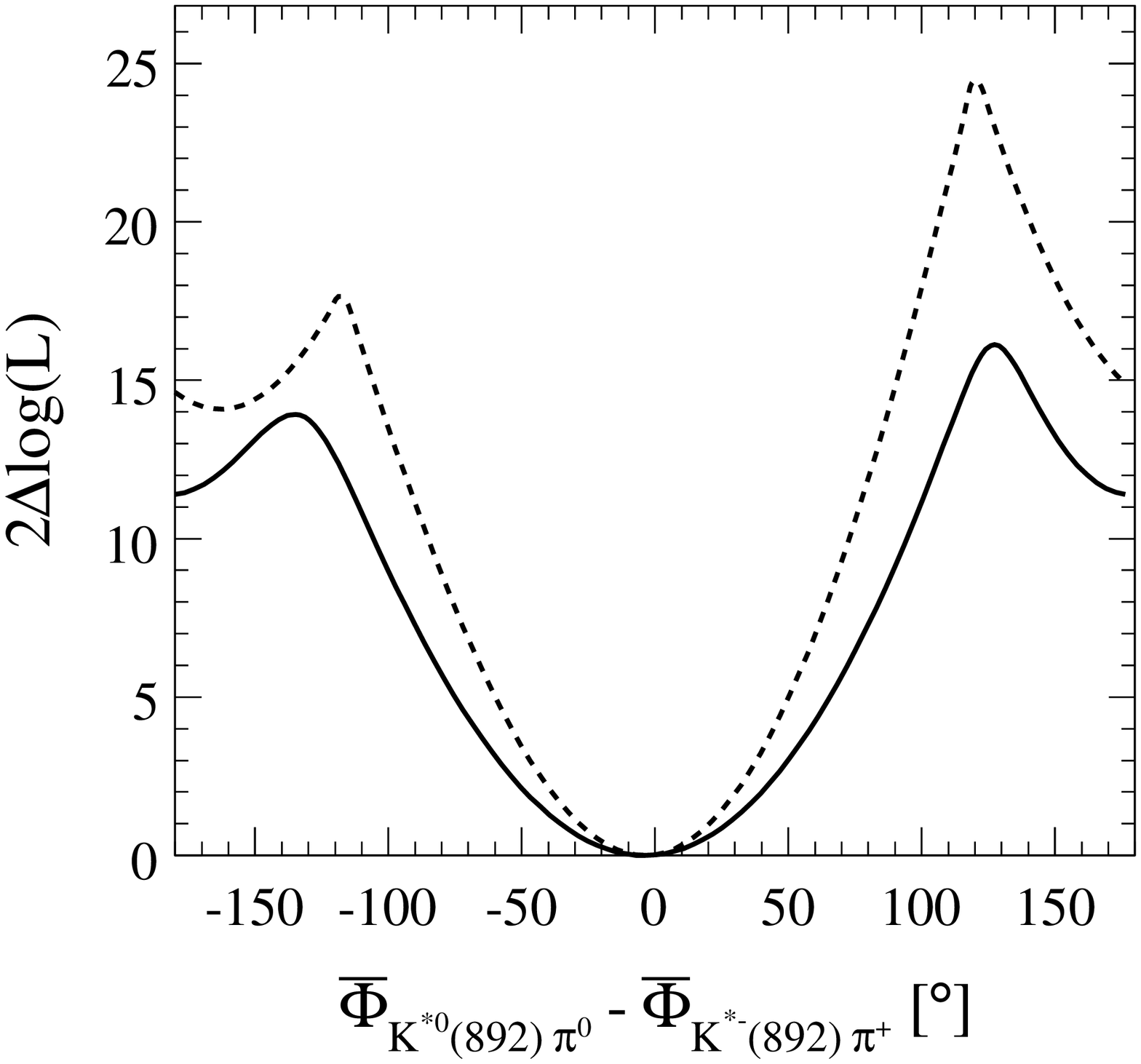,width=7cm}  
    \caption{\label{fig:kstarpidiff} Likelihood scans illustrating the measurements of $\Phi_{\Kstarz\piz} - \Phi_{\Kstarp\pim} = -22\pm 24~({\rm stat.})\pm 18~({\rm syst.})^{\circ}$ (top) and $\overline{\Phi}_{\Kstarz\piz} - \overline{\Phi}_{\Kstarm\pip} = -4\pm 24~({\rm stat.})\pm 19~({\rm syst.})^{\circ}$ (bottom). The solid (dashed) line shows the $2\Delta{\rm log}(L)$ for the total (statistical) uncertainty.}
  \end{center}
\end{figure}

\noindent Likelihood scans illustrating the measurements of $\PhiorPhibar_{\Kstarz\piz} - \PhiorPhibar_{\Kstarp\pim}$ are shown in~\figref{kstarpidiff}. We measure $\Phi_{\Kstarz\piz} - \Phi_{\Kstarp\pim} = -22\pm 24~({\rm stat.})\pm 18~({\rm syst.})^{\circ}$ and $\overline{\Phi}_{\Kstarz\piz} - \overline{\Phi}_{\Kstarm\pip} = -4\pm 24~({\rm stat.})\pm 19~({\rm syst.})^{\circ}$ using the helicity convention defined in~\figref{Helicity}. We use $\Delta{\phi_{\Kstar\pi}} = 58\pm 33~({\rm stat.})\pm 9~({\rm syst.})^{\circ}$~\cite{babar-kspipi} and subtract the mixing phase contribution $2\beta=42.2\pm1.8^{\circ}$~\cite{HFAG} to evaluate~\equaref{Phi3half}.

\begin{figure}[h!]
\epsfig{file=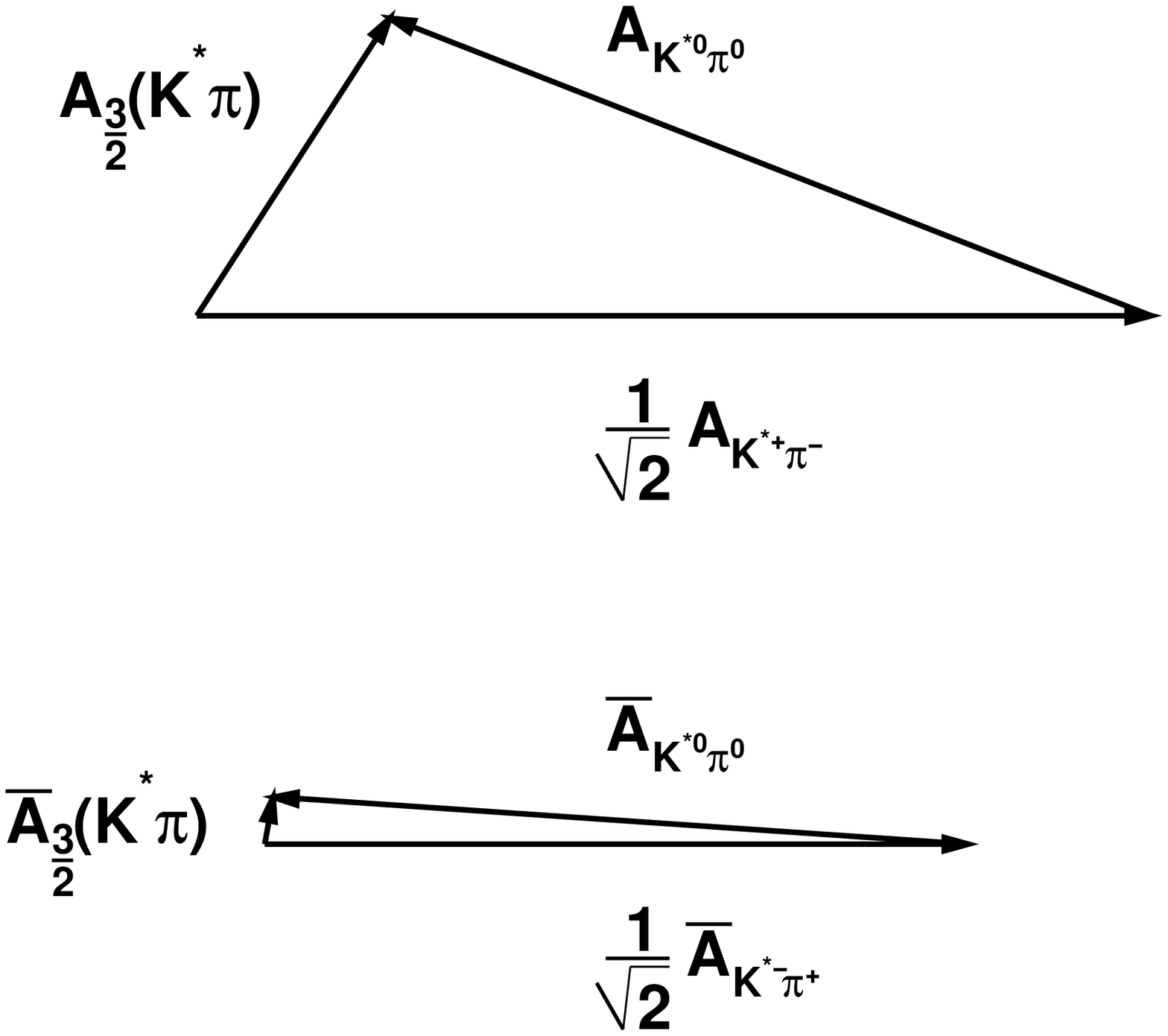,width=7.5cm}
\caption{\label{fig:A32_Kpi} Isospin triangles drawn to scale for $\B\to\Kstar\pi$ decays. The isobar amplitudes are summarized in~\tabref{fourfitsolutions} as solution I. Note that the isospin triangle for $\Bbar$ decays is relatively flat and ${\cal\Abar}_{3\over2}(\Kstar\pi)$ is consistent with 0.}
\end{figure}

It is important to note that for vector resonances the helicity convention defines an ordering of particles in the final state via the angular dependence $T(J,x,y)$~(\tabref{Zblatt}). This means that care must be taken to use a consistent helicity convention when evaluating an isospin decomposition of vector amplitudes~\cite{Gronau:helicity}. In this analysis the helicity angle for $\Kstarp\pim$ is measured between the $\Kp$ and $\pim$ while the helicity angle for $\Kstarz\piz$ is measured between the $\pim$ and $\piz$. This results in a sign flip between the $\AorAbar_{\Kstarz\piz}$, $\AorAbar_{\Kstarp\pim}$ amplitudes when~\equaref{A32} is evaluated with $\Kstar(892)\pi$ amplitudes measured as in~\tabref{bestsolution}. The $\AorAbar_{3\over2}(\Kstar\pi)$ isospin triangles described by~\equaref{A32} are shown in~\figref{A32_Kpi} for the $\Kstar(892)\pi$ amplitudes measured in~\tabref{bestsolution}. The destructive interference between $\Kstar(892)\pi$ amplitudes in the isospin decomposition~(\figref{A32_Kpi}) is expected, since these amplitudes are penguin-dominated while $\AorAbar_{3\over2}$ is penguin-free by construction~\cite{Gronau:helicity}. We find that ${\cal\Abar}_{3\over2}(\Kstar\pi)$ is consistent with 0 and consequently that the uncertainty in $\overline{\delta}$ is too large to permit a measurement of $\Phi_{3\over2}$ using $\Kstar(892)\pi$ amplitudes as originally suggested in Ref.~\cite{Ciuchini:2006kv}. 

\subsubsection{$\B\to\rho K$ decays}

It is also possible to obtain a CKM constraint using $\B\to\rho K$ decay amplitudes as in~\equaref{A32_rhoK}. Here, the $\rho^-\Kp$ and $\rho^0\Kz$ amplitudes do not decay to a common final state, making a direct measurement of their relative phase impossible. Interference between $\rho^-\Kp$ and $\rho^0\Kz$ amplitudes and $\Kstarp\pim$ can be observed using $\Bz$ decays to both $\Kp\pim\piz$ and $\KS\pip\pim$, permitting an indirect measurement of their relative phase. The weak phase of ${\cal A}_{3\over2}(\rho K)$ is given by~\equaref{Phi3half} where now, 

\begin{equation}
\label{eq:phi3half_rhoK}
\deltaordeltabar = \arctan \Bigg({{|\AorAbar_{\rho^-\Kp}|\sorsbar_- + \sqrt{2}|\AorAbar_{\rho^0\Kz}|\sorsbar_0}\over{|\AorAbar_{\rho^-\Kp}|\corcbar_- + \sqrt{2}|\AorAbar_{\rho^0\Kz}|\corcbar_0}} \Bigg).
\end{equation}

\noindent Here we define 

\begin{eqnarray}
\sorsbar_- &=& \sin{(\PhiorPhibar_{\rho^-\Kp} - \PhiorPhibar_{\Kstarp\pim})}, \\
\corcbar_- &=& \cos{(\PhiorPhibar_{\rho^-\Kp} - \PhiorPhibar_{\Kstarp\pim})}, \\ 
\sorsbar_0 &=& \sin{(\PhiorPhibar_{\rho^0\Kz} - \PhiorPhibar_{\Kstarp\pim})}, \\
\corcbar_0 &=& \cos{(\PhiorPhibar_{\rho^0\Kz} - \PhiorPhibar_{\Kstarp\pim})}. 
\end{eqnarray}

\begin{figure}[h]
  \begin{center}
    \epsfig{file=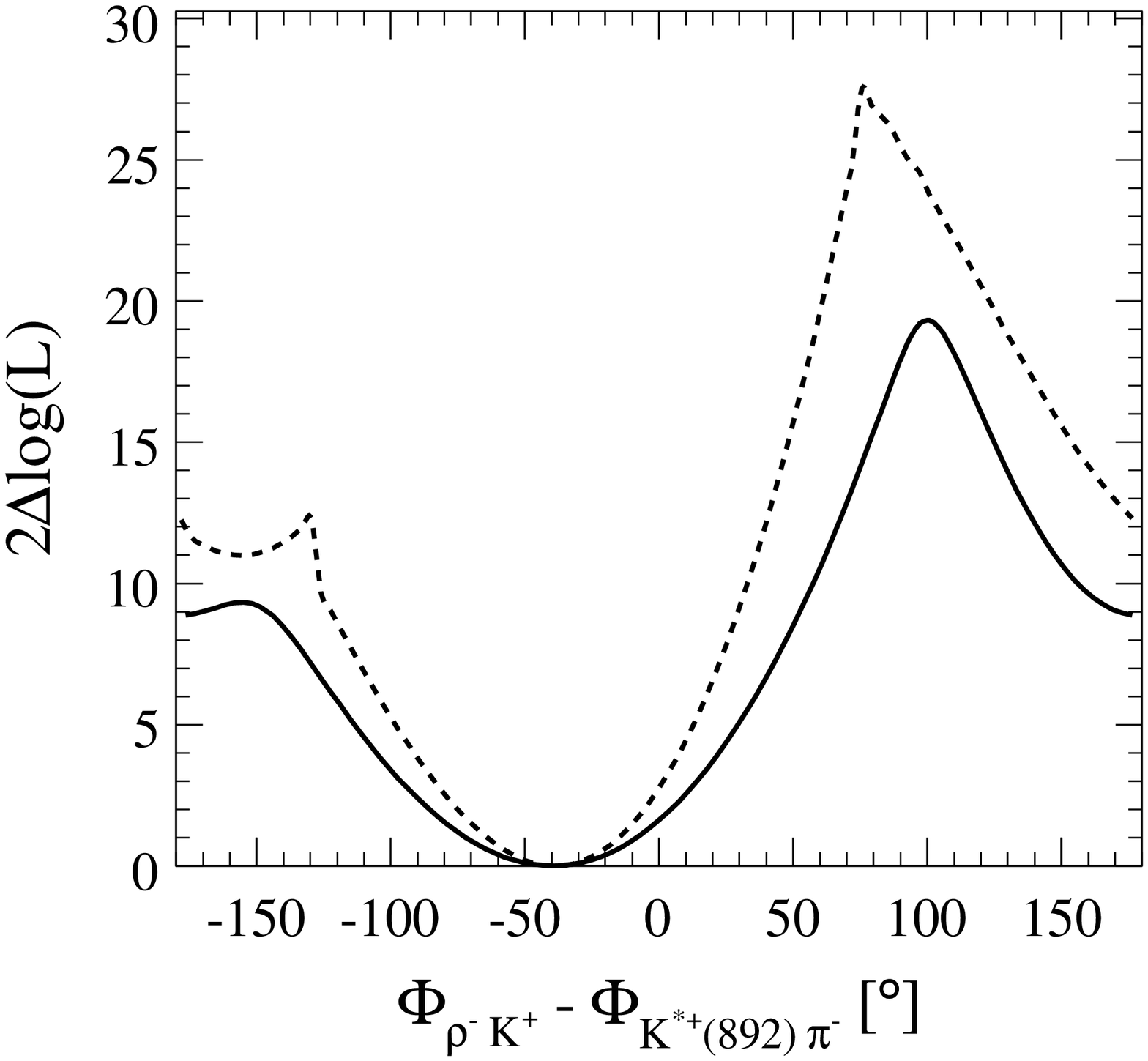,width=7cm} 
    \epsfig{file=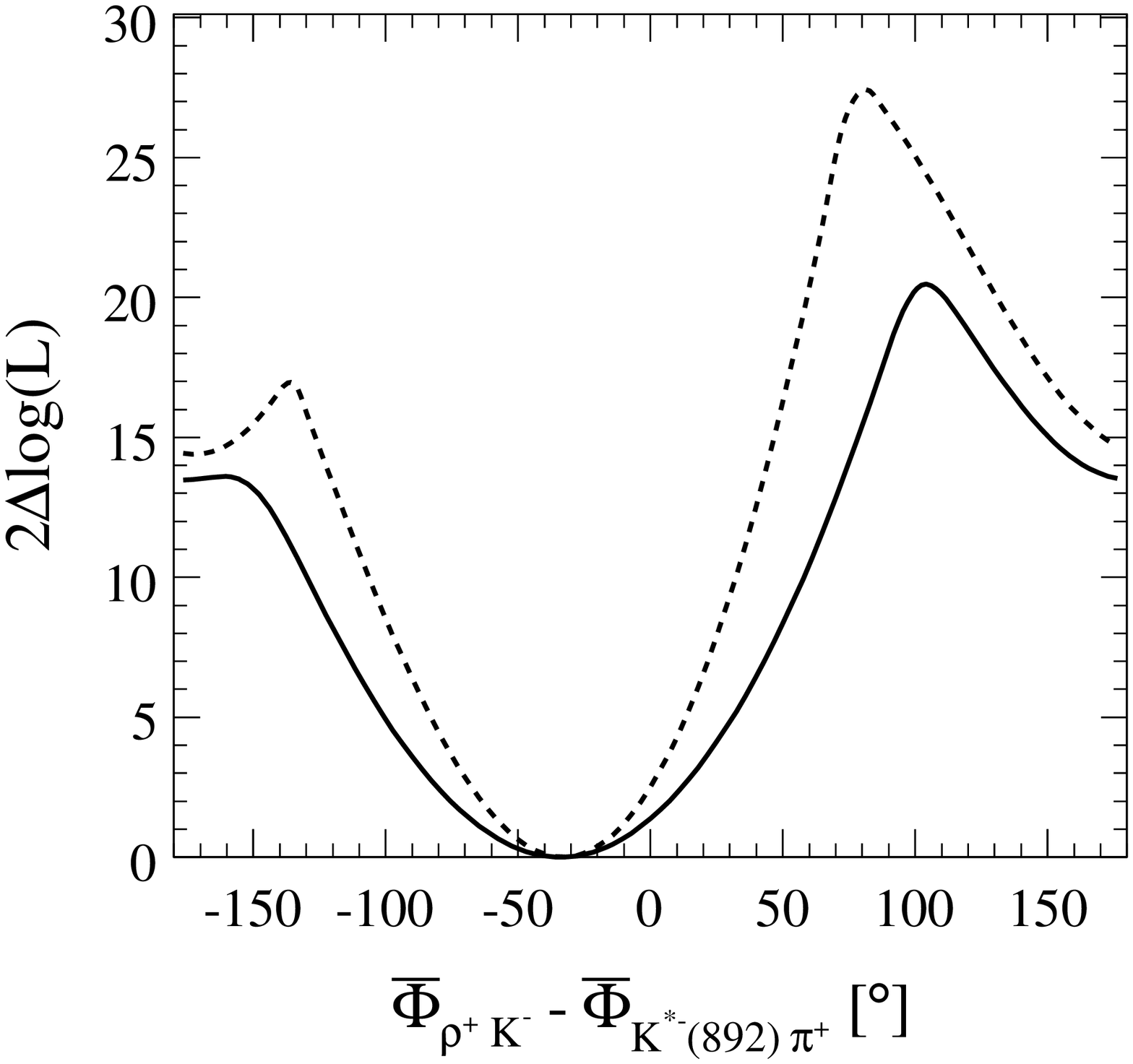,width=7cm}   
    \caption{\label{fig:rhoKdiff} Likelihood scans illustrating the measurements of $\Phi_{\rho^-\Kp} - \Phi_{\Kstarp\pim}=-33\pm22~({\rm stat.}) \pm 20~({\rm syst.})^\circ$ (top) and $\overline{\Phi}_{\rho^+\Km} - \overline{\Phi}_{\Kstarm\pip}=-39\pm25~({\rm stat.}) \pm 20~({\rm syst.})^\circ$ (bottom). The solid (dashed) line shows the $2\Delta{\rm log}(L)$ for the total (statistical) uncertainty.}
  \end{center}
\end{figure}

\noindent We use the $\Kstar(892)^+\pim$ amplitude in the evaluation of these expressions. Likelihood scans of the phase differences $\PhiorPhibar_{\rho^-\Kp} - \PhiorPhibar_{\Kstarp\pim}$ are shown in~\figref{rhoKdiff}. We measure $\Phi_{\rho^-\Kp} - \Phi_{\Kstarp\pim}=-33\pm22~({\rm stat.}) \pm 20~({\rm syst.})^\circ$ and $\overline{\Phi}_{\rho^+\Km} - \overline{\Phi}_{\Kstarm\pip}=-39\pm25~({\rm stat.}) \pm 20~({\rm syst.})^\circ$.

\noindent We use the measurements $\Phi_{\rho^0\Kz} - \Phi_{\Kstarp\pim} = 174\pm 28~({\rm stat.})\pm 15~({\rm syst.})^\circ$ and $\overline{\Phi}_{\rho^0\Kz} - \overline{\Phi}_{\Kstarm\pip} = -148\pm 25~({\rm stat.}) \pm 16~({\rm syst.})^\circ$ given in Ref.~\cite{babar-kspipi}. Before evaluating~\equaref{phi3half_rhoK} we must account for any discrepancy in helicity conventions used in this analysis and Ref.~\cite{babar-kspipi}. Here we must consider the helicity conventions used not only by the $\rho K$ amplitudes but also the intermediate $\Kstar(892)^+\pim$ amplitude. In this analysis the helicity angle is measured between the $\piz$ and $\Kp$ for $\rho^-\Kp$ amplitudes while the helicity angle is measured between the $\pim$ and $\KS$ in Ref.~\cite{babar-kspipi}. It is also the case that the helicity angle is measured between the $\pip$ and $\pim$ for $\Kstarp$ decays in Ref.~\cite{babar-kspipi}, and is measured between the $\Kp$ and $\pim$ in this analysis. Since there are a total of two sign flips due to these differences there is no net sign flip between ${\cal A}_{\rho^0\Kz}$ and ${\cal A}_{\rho^-\Kp}$ when~\equaref{A32_rhoK} is evaluated using the measurements presented in this article and in Ref.~\cite{babar-kspipi}.  

\begin{figure}[h]
  \epsfig{file=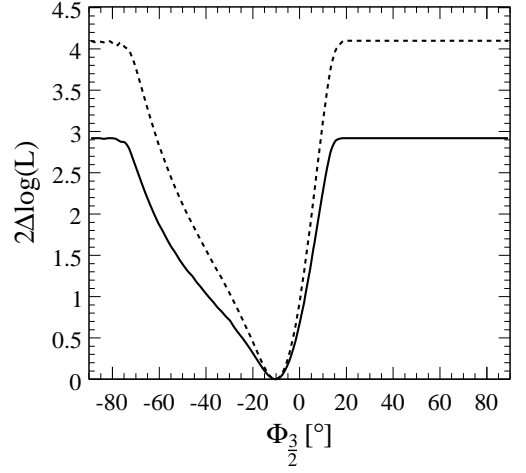,width=7cm}   
  \caption{\label{fig:rhoK_phi32} Likelihood scans illustrating the measurement of $\Phi_{3\over2}=-10^{+10}_{-20}~({\rm stat.})^{+7}_{-22}~({\rm syst.})^\circ$. The solid (dashed) line shows the $2\Delta{\rm log}(L)$ for the total (statistical) uncertainty.}
\end{figure} 

\begin{figure}[h]
  \epsfig{file=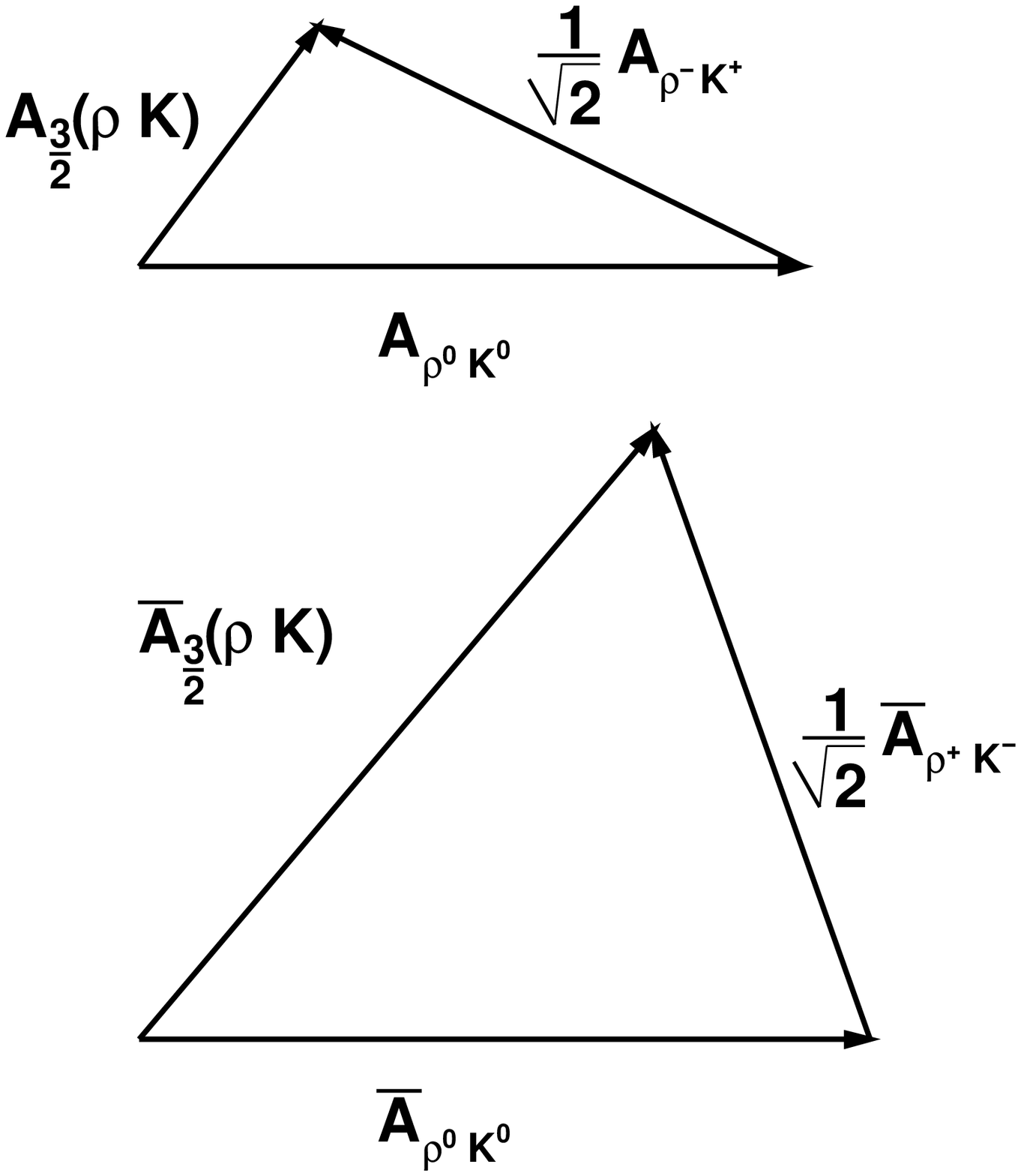,width=7cm}
  \caption{\label{fig:A32_RhoK} Isospin triangles drawn to scale for $\B\to\rho K$ decays. The isobar amplitudes are summarized in~\tabref{fourfitsolutions} as solution I and in Ref.~\cite{babar-kspipi}.}
\end{figure}

We evaluate~\equaref{Phi3half} using~\equaref{phi3half_rhoK} and produce a measurement of $\Phi_{3\over2}$ illustrated in~\figref{rhoK_phi32}. The $\AorAbar_{3\over2}(\rho K)$ isospin triangles described by~\equaref{A32_rhoK} are shown in~\figref{A32_RhoK} for the $\B\to\rho K$ amplitudes measured in this analysis and Ref.~\cite{babar-kspipi}. In contrast to the $\Kstar\pi$ isospin triangles~(\figref{A32_Kpi}) both $\AorAbar_{3\over2}(\rho K)$ are significantly different from 0 permitting a measurement of $\Phi_{3\over2}=-10^{+10}_{-20}~({\rm stat.})^{+7}_{-22}~({\rm syst.})^\circ$. This measurement is defined modulo $180^\circ$ (see~\equaref{Phi3half}) and we quote only the value between $\pm 90^\circ$. The likelihood constraint shown in~\figref{rhoK_phi32} becomes flat, since sufficiently large deviations of the $\B\to\rho K$ amplitudes will result in a flat isospin triangle and consequently an arbitrary value of $\Phi_{3\over2}$. 

\subsection{Evaluation of the amplitude sum rule}
\label{sec:sumrule}

The sum rule given in~\equaref{sumrule} motivates the definition of the dimensionless quantity, 

\begin{equation}
\label{eq:Sigma32}
\Sigma_{3\over2} = {{|\overline{{\cal A}}_{3 \over 2}(\Kstar\pi)|^2 - |{\cal A}_{3 \over 2}(\Kstar\pi)|^2}\over{|\overline{{\cal A}}_{3 \over 2}(\rho K)|^2 - |{\cal A}_{3 \over 2}(\rho K)|^2}} + 1. 
\end{equation}

\begin{figure}[h]
  \epsfig{file=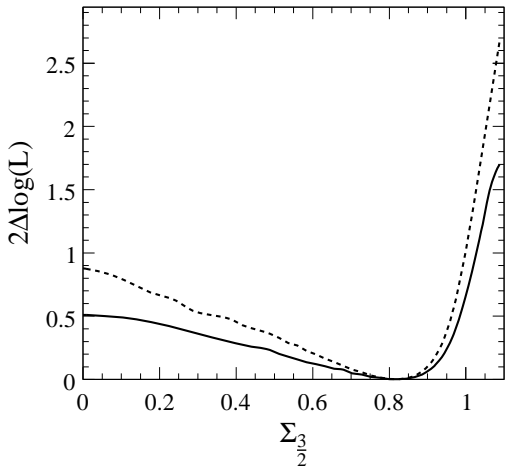,width=7cm}
  \caption{\label{fig:Sigma32} Likelihood scan for $\Sigma_{3\over2}$. The solid (dashed) line shows the $2\Delta{\rm log}(L)$ for the total (statistical) uncertainty. We measure $\Sigma_{3\over2}=0.82^{+0.18}_{-0.92}~({\rm stat.})^{+0.11}_{-1.35}~({\rm syst.})$.}
\end{figure}

\noindent The asymmetry parameter $\Sigma_{3\over2}$ will be 0 in the limit of exact SU(3) symmetry. Deviations from exact SU(3) symmetry or contributions from new physics operators can be quantified, if $\Sigma_{3\over2}$ is measured to be significantly different from 0. We use the amplitudes and phase differences among $\Bz\to\Kstar\pi$ and $\Bz\to\rho K$ amplitudes as described in~\secref{Phi32} to produce a likelihood scan of $\Sigma_{3\over2}$ as shown in~\figref{Sigma32}. We measure $\Sigma_{3\over2}=0.82^{+0.18}_{-0.92}~({\rm stat.})^{+0.11}_{-1.35}~({\rm syst.})$, consistent with 0. The large statistical and systematic uncertainties are due to the poorly measured phase differences between the $\rho K$ and $\Kstar\pi$ amplitudes.  

\subsection{Evidence of direct \CP~violation in $\bf \Bz\to\Kstarp\pim$ decays}
\label{sec:acp}

\begin{figure}[h]
  \epsfig{file=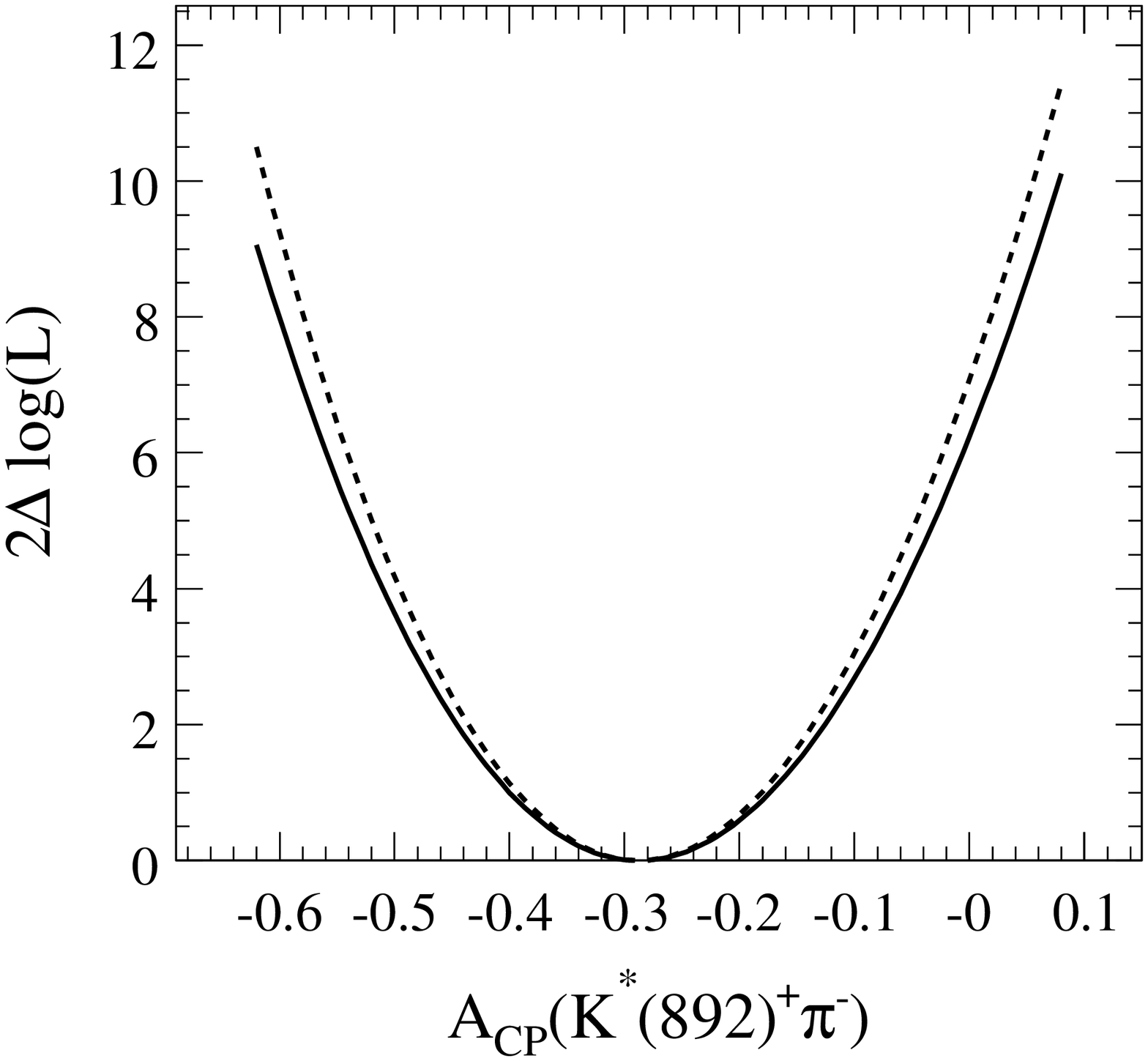,width=7cm} 
  \epsfig{file=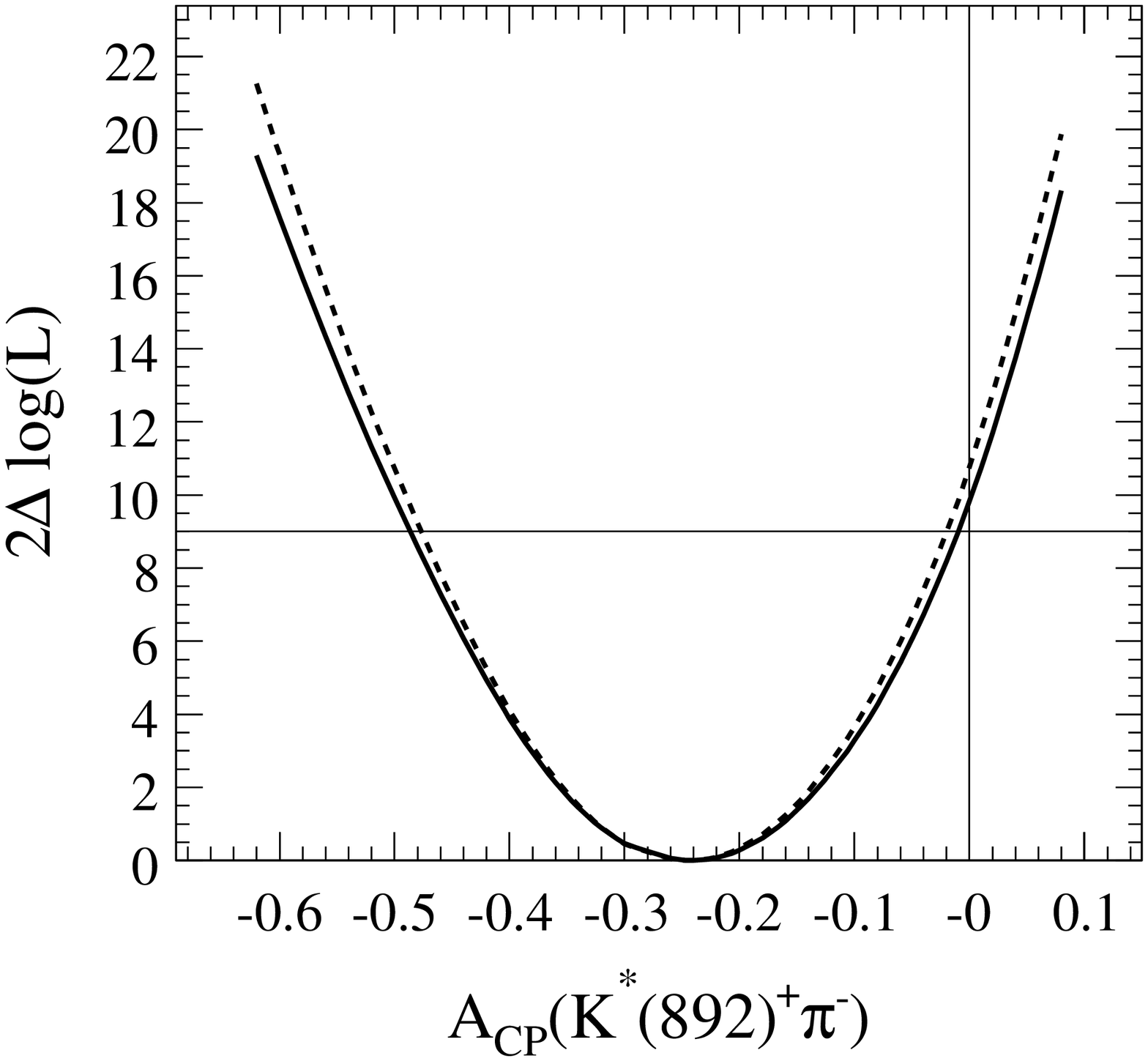,width=7cm} 
  \caption{\label{fig:Acpscans} Likelihood scans for $A_{C\!P}(\Kstar(892)^+\pim)$ using only the $\Bz\to\Kp\pim\piz$ analysis (top) and the combined measurement with the $\Bz\to \KS\pip\pim$ analysis (bottom). The solid (dashed) line shows the $2\Delta{\rm log}(L)$ for the total (statistical) uncertainty. We measure $A_{C\!P}(\Kstarp\pim)= -0.29 \pm 0.11~({\rm stat.})\pm 0.02~({\rm syst.})$~(\tabref{bestsolution}) in $\Bz\to\Kp\pim\piz$ and $A_{C\!P}(\Kstarp\pim)=-0.24 \pm 0.07~({\rm stat.}) \pm 0.02~({\rm syst.})$ when the measurement in $\Bz\to\KS\pip\pim$ is combined. The vertical line highlights $A_{C\!P}=0$ and the horizontal line corresponds to $2\Delta{\rm log}{\cal L}=9$, i.e. $3~\sigma$.}
\end{figure}

Measurements of direct \CP~violation are made in both the analyses of $\Bz\to\Kp\pim\piz$ and $\Bz\to \KS\pip\pim$. Since these analyses are statistically independent, the measurements of $A_{C\!P}$ may be combined for intermediate resonances common to both. The combined measurement of direct \CP~violation for $\Bz\to\Kstar(892)^+\pim$ decays is found to be $A_{C\!P}(\Kstarp\pim) = -0.24 \pm 0.07~({\rm stat.}) \pm 0.02~({\rm syst.})$ and is significant at $3.1~\sigma$. Likelihood scans illustrating the measurement of $A_{C\!P}(\Kstarp\pim)$ in $\Bz\to\Kp\pim\piz$ and the combined result including the measurement in $\Bz\to \KS\pip\pim$ are shown in~\figref{Acpscans}.  

\section{SUMMARY}
\label{sec:Summary}

In summary, we analyze the DP distribution for $\Bz\to\Kp\pim\piz$ decays from a sample of $454$ million $\BB$ pairs. We determine branching fractions, \CP~asymmetries and phase differences of seven intermediate resonances in addition to a NR contribution. We find that the isospin amplitude constructed from $\Bz\to\Kstar\pi$ amplitudes is consistent with 0, preventing the possibility of a useful CKM constraint as originally suggested in Ref.~\cite{Ciuchini:2006kv}. A similar construction made with $\Bz\to\rho K$ amplitudes provides sufficient sensitivity to measure the weak phase of the isospin amplitude. We measure $\Phi_{3\over2}=-10^{+10}_{-20}~({\rm stat.})^{+7}_{-22}~({\rm syst.})^\circ$ using $\Bz\to\rho K$ amplitudes. Fundamentally, the sensitivity of $\Bz\to\Kstar\pi$ and $\Bz\to\rho K$ decay amplitudes to the CKM angle $\gamma$ is limited by their QCD penguin dominance~\cite{Gronau:helicity}, the isopin amplitude constructed from a linear combination of such amplitudes being QCD penguin free. We suggest that isospin combinations of $B_s\to\Kstar\pi$ and $B_s\to\rho K$ amplitudes, which are not QCD penguin dominated, will provide a much more sensitive CKM constraint~\cite{Ciuchini2007201}. We also produce the first test of a \CP~rate asymmetry sum rule (\equaref{sumrule}) using isospin amplitudes. We find the violation of this sum rule to be $\Sigma_{3\over2}=0.82^{+0.18}_{-0.92}~({\rm stat.})^{+0.11}_{-1.35}~({\rm syst.})$, consistent with 0. A significant violation of the sum rule could indicate the presence of new physics operators~\cite{Gronau:sumrule}, making further study of the isospin amplitudes presented in this paper an interesting area of study. Finally, we find evidence of direct \CP~violation in $\Bz\to\Kstarp\pim$ decays shown in~\figref{Acpscans}, $A_{C\!P}(\Kstar(892)^+\pim) = -0.24 \pm 0.07~({\rm stat.}) \pm 0.02~({\rm syst.})$, when measurements from the $\Bz\to\Kp\pim\piz$ and $\Bz\to\KS\pip\pim$~\cite{babar-kspipi} DP analyses are combined.

\section{Acknowledgments}
\label{sec:acknowledgments}
We thank Michael Gronau, Dan Pirjol and Jonathan Rosner for useful discussions.
We are grateful for the 
extraordinary contributions of our \pep2\ colleagues in
achieving the excellent luminosity and machine conditions
that have made this work possible.
The success of this project also relies critically on the 
expertise and dedication of the computing organizations that 
support \babar. 
The collaborating institutions wish to thank 
SLAC for its support and the kind hospitality extended to them. 
This work is supported by the
US Department of Energy
and National Science Foundation, the
Natural Sciences and Engineering Research Council (Canada),
the Commissariat \`a l'Energie Atomique and
Institut National de Physique Nucl\'eaire et de Physique des Particules
(France), the
Bundesministerium f\"ur Bildung und Forschung and
Deutsche Forschungsgemeinschaft
(Germany), the
Istituto Nazionale di Fisica Nucleare (Italy),
the Foundation for Fundamental Research on Matter (The Netherlands),
the Research Council of Norway, the
Ministry of Education and Science of the Russian Federation, 
Ministerio de Ciencia e Innovaci\'on (Spain), and the
Science and Technology Facilities Council (United Kingdom).
Individuals have received support from 
the Marie-Curie IEF program (European Union), the A. P. Sloan Foundation (USA) 
and the Binational Science Foundation (USA-Israel).

\bibliography{note2291}
\bibliographystyle{apsrev}
\clearpage
\section*{APPENDIX}
\label{sec:appendix}

The results of the four solutions found in the ML fit are summarized in~\tabref{fourfitsolutions}. Only the statistical uncertainties are quoted in this summary. The systematic uncertainties are summarized in Tables~\ref{tab:systTable_K} and~\ref{tab:systTable_rho}. The \CP~averaged interference fractions, $I_{kl}$, among the intermediate decay amplitudes are given in~\tabref{IF} for solution I, expressed as a percentage of the total charmless decay amplitude. Here,

\begin{equation}
I_{kl}=\frac{|{\cal A}_k|^2 + |\overline{\cal A}_k|^2}{|\sum\limits_{j}{\cal A}_j|^2+|\sum\limits_{j}\overline{\cal A}_j|^2},
\end{equation} 

\noindent for $k=l$ and, 

\begin{equation}
I_{kl}=\frac{{\cal A}_k{\cal A}_l^* + {\cal A}_k^*{\cal A}_l + \overline{\cal A}_k\overline{\cal A}_l^* + \overline{\cal A}_k^*\overline{\cal A}_l}{|\sum\limits_{j}{\cal A}_j|^2+|\sum\limits_{j}\overline{\cal A}_j|^2}
\end{equation} 

\noindent for $k\neq l$. The ${\cal A}_k$ are the amplitudes defined in~\equaref{isobars}. 

The full correlation matrix for solution I is given in Tables~\ref{tab:Corr_Amp_I},~\ref{tab:Corr_Amp_II},~\ref{tab:Corr_Ampbar_I}, and~\ref{tab:Corr_Ampbar_II}. The tables are separated by correlations among $\B$ and $\Bbar$ decay amplitudes.

\begin{table*}[h!]
\begin{center}
\caption{Summary of fit results for the four solutions. The isobar parameters $\protect\aorabar$ and $\protect\PhiorPhibar$ are defined in~\equaref{isobars}. The phases $\protect\PhiorPhibar$ are measured relative to $\Bz(\Bzb)\to\rho (770)^\mp K^\pm$ in degrees and the $\protect\aorabar$ are measured relative to $\Bzb\to\rho (770)^+ K^-$ so that $\overline{a}_{\rho(770)^+ K^-}=1$. The uncertainties are statistical only.}
\label{tab:fourfitsolutions}
\setlength{\tabcolsep}{1.2pc}
\begin{tabular}{cccccc}
\hline\hline

Amplitude  &  Parameter                   &  Solution-I         &  Solution-II          & Solution-III        & Solution-IV  \\ 
           &  $\Delta{\rm NLL}$           &  0.00               &  5.43                 & 7.04                & 12.33 \\ 
\hline
\multirow{4}{*}{$\rho(770)^- K^+$} &  $a$ 	          &  0.82 $\pm$ 0.08    &  0.82 $\pm$ 0.09      & 0.83 $\pm$ 0.07     & 0.84 $\pm$ 0.10 \\
                  & $\overline{a}$        &  1 (fixed)          &  1 (fixed)            & 1 (fixed)           & 1 (fixed) \\
                  & $\Phi$  	          &  0 (fixed)	        &  0 (fixed)	        & 0 (fixed)           & 0 (fixed) \\
                  & $\overline{\Phi}$  	  &  0 (fixed)	        &  0 (fixed)	        & 0 (fixed)           & 0 (fixed)\\

\hline 
\multirow{4}{*}{$\rho(1450)^- K^+$} &  $a$                 &  0.57 $\pm$ 0.14	&  0.48 $\pm$ 0.26	&  0.59 $\pm$ 0.12      & 0.49 $\pm$ 0.20 \\
                   & $\overline{a}$       &  0.52 $\pm$ 0.15	&  0.52 $\pm$ 0.16	&  0.54 $\pm$ 0.13      & 0.55 $\pm$ 0.22 \\
                   & $\Phi$  		  &  126 $\pm$ 25	&  90 $\pm$ 22         	&  126 $\pm$ 25         & 89 $\pm$ 22 \\
                   & $\overline{\Phi}$    &  74 $\pm$ 19	&  74 $\pm$ 18         	&  72 $\pm$ 20          & 71 $\pm$ 21 \\

\hline
\multirow{4}{*}{$\rho(1700)^- K^+$} &  $a$  		  &  0.33 $\pm$ 0.15	&  0.11 $\pm$ 0.31	&  0.34 $\pm$ 0.13      & 0.11 $\pm$ 0.31  \\
                   & $\overline{a}$  	  &  0.23 $\pm$ 0.12	&  0.23 $\pm$ 0.12	&  0.17 $\pm$ 0.12      & 0.17 $\pm$ 0.17  \\
                   & $\Phi$  		  &  50 $\pm$ 38	&  35 $\pm$ 164        	&  50 $\pm$ 34          & 34 $\pm$ 159  \\
                   & $\overline{\Phi}$    &  18 $\pm$ 36	&  17 $\pm$ 35         	&  -15 $\pm$ 48         & -17 $\pm$ 57 \\

\hline
\multirow{4}{*}{$\Kstar(892)^+\pi^-$} &  $a$  		  &  0.66 $\pm$ 0.06	&  0.66 $\pm$ 0.07	&  0.67 $\pm$ 0.05      & 0.68 $\pm$ 0.08   \\
                   & $\overline{a}$       &  0.49 $\pm$ 0.06	&  0.49 $\pm$ 0.06	&  0.55 $\pm$ 0.06      & 0.54 $\pm$ 0.08  \\
                   & $\Phi$  		  &  39 $\pm$ 25	&  156 $\pm$ 25        	&  40 $\pm$ 25          & 156 $\pm$ 25  \\
                   & $\overline{\Phi}$    &  33 $\pm$ 22	&  33 $\pm$ 22         	&  172 $\pm$ 20         & 172 $\pm$ 21  \\

\hline
\multirow{4}{*}{$\Kstar(892)^0\pi^0$} &  $a$  		  &  0.57 $\pm$ 0.06	&  0.57 $\pm$ 0.06	&  0.58 $\pm$ 0.06      & 0.58 $\pm$ 0.07  \\
    		   & $\overline{a}$  	  &  0.49 $\pm$ 0.05	&  0.49 $\pm$ 0.05	&  0.50 $\pm$ 0.05      & 0.51 $\pm$ 0.07  \\
   		   & $\Phi$ 		  &  17 $\pm$ 20	&  17 $\pm$ 21         	&  17 $\pm$ 21          & 16 $\pm$ 21  \\
		   & $\overline{\Phi}$    &  29 $\pm$ 18	&  29 $\pm$ 18         	&  9 $\pm$ 18           & 9 $\pm$ 19  \\

\hline
\multirow{4}{*}{$(K\pi)^{*+}_0\pi^-$} &  $a$     	  &  1.15 $\pm$ 0.09	&  1.22 $\pm$ 0.10	&  1.18 $\pm$ 0.07     & 1.24 $\pm$ 0.13  \\
  		     & $\overline{a}$  	  &  1.24 $\pm$ 0.09	&  1.24 $\pm$ 0.09	&  1.32 $\pm$ 0.08     & 1.33 $\pm$ 0.14  \\
  		     & $\Phi$ 		  &  -130 $\pm$ 22	&  -19 $\pm$ 25	        &  -130 $\pm$ 22        & -19 $\pm$ 24  \\
  		     & $\overline{\Phi}$  &  -167 $\pm$ 16	&  -168 $\pm$ 16	&  -38 $\pm$ 18         & -38 $\pm$ 19  \\

\hline
\multirow{4}{*}{$(K\pi)^{*0}_0\pi^0$} &  $a$     	  &  0.91 $\pm$ 0.07 	&  1.25 $\pm$ 0.10	&  0.93 $\pm$ 0.07     & 1.28 $\pm$ 0.13  \\
  		     & $\overline{a}$  	  &  0.78 $\pm$ 0.08	&  0.78 $\pm$ 0.09	&  1.11 $\pm$ 0.07     & 1.12 $\pm$ 0.11  \\
   		     & $\Phi$ 		  &  10 $\pm$ 17	&  21 $\pm$ 17	        &  10 $\pm$ 17          & 21 $\pm$ 17  \\
  		     & $\overline{\Phi}$  &  13 $\pm$ 17	&  13 $\pm$ 17	        &  1 $\pm$ 14           & 1 $\pm$ 14 \\

\hline
\multirow{4}{*}{NR}             &  $a$     	  &  0.56 $\pm$ 0.08	&  0.31 $\pm$ 0.09	&  0.58 $\pm$ 0.08    & 0.32 $\pm$ 0.10  \\
    		     & $\overline{a}$  	  &  0.62 $\pm$ 0.07	&  0.63 $\pm$ 0.07	&  0.57 $\pm$ 0.08     & 0.58 $\pm$ 0.09  \\
    		     & $\Phi$  		  &  87 $\pm$ 21	&  -61 $\pm$ 22	        &  87 $\pm$ 21          & -61 $\pm$ 22  \\
  		     & $\overline{\Phi}$  &  48 $\pm$ 14	&  48 $\pm$ 14	        &  -65 $\pm$ 15         & -65 $\pm$ 17  \\
\hline
 
\hline\hline
\end{tabular}
\end{center}
\end{table*}

\begin{table*}[h!]
\begin{center}
\caption{\label{tab:systTable_K} Systematic uncertainties associated with the $\Bz\to\Kstar\pi$ isobar parameters summarized in~\tabref{fourfitsolutions} under Sol. I. Uncertainties in the phases are in degrees.}
\setlength{\tabcolsep}{.9pc}
{\begin{tabular}{cccccc}
\hline\hline
Amplitude  & &  $\sigma (a)$ &  $\sigma (\overline{a})$   &  $\sigma (\Phi)~[^\circ]$ &  $\sigma (\overline{\Phi})~[^\circ]$ \\
\hline
\multirow{9}{*}{$\Kstar(892)^+\pim$}
 & Isobar Model         & 0.00 & 0.00 & 16 & 19\\
 & \B Backgrounds        & 0.01 & 0.01 & 1 & 1\\
 & PDF Shape Parameters & 0.03 & 0.01 & 3 & 1\\
 & SCF Fraction         & 0.00 & 0.00 & 1 & 1\\
 & PID Systematics      & 0.00 & 0.00 & 0 & 1\\
 & Lineshapes           & 0.01 & 0.00 & 9 & 4\\
 & Fit Bias             & 0.02 & 0.01 & 6 & 6\\
 & Continuum DP         & 0.00 & 0.00 & 2 & 1\\
 & {\bf Total}          & {\bf 0.04} & {\bf 0.02} & {\bf 20} & {\bf 20}\\                                 
\hline
\multirow{9}{*}{$(K\pi)^{*+}_0\pim$}
 & Isobar Model         & 0.02 & 0.02 & 19 & 36\\
 & \B Backgrounds        & 0.01 & 0.01 & 1 & 1\\
 & PDF Shape Parameters & 0.04 & 0.05 & 3 & 1\\
 & SCF Fraction         & 0.00 & 0.00 & 1 & 1\\
 & PID Systematics      & 0.01 & 0.01 & 0 & 1\\
 & Lineshapes           & 0.01 & 0.01 & 8 & 3\\
 & Fit Bias             & 0.03 & 0.03 & 6 & 6\\
 & Continuum DP         & 0.03 & 0.03 & 4 & 5\\
 & {\bf Total}          & {\bf 0.06} & {\bf 0.07} & {\bf 22} & {\bf 37}\\                                  
\hline
\multirow{9}{*}{$\Kstar(892)^0\piz$}
 & Isobar Model         & 0.02 & 0.01 & 2 & 0\\
 & \B Backgrounds        & 0.01 & 0.01 & 1 & 1\\
 & PDF Shape Parameters & 0.02 & 0.02 & 1 & 1\\
 & SCF Fraction         & 0.00 & 0.00 & 0 & 0\\
 & PID Systematics      & 0.00 & 0.00 & 1 & 0\\
 & Lineshapes           & 0.01 & 0.00 & 4 & 4\\
 & Fit Bias             & 0.01 & 0.00 & 1 & 1\\
 & Continuum DP         & 0.01 & 0.01 & 6 & 5\\
 & {\bf Total}          & {\bf 0.03} & {\bf 0.02} & {\bf 8} & {\bf 6}\\                                   
\hline
\multirow{9}{*}{$(K\pi)^{*0}_0\piz$}
 & Isobar Model         & 0.02 & 0.03 & 14 & 9\\
 & \B Backgrounds        & 0.01 & 0.02 & 1 & 1\\
 & PDF Shape Parameters & 0.03 & 0.02 & 1 & 1\\
 & SCF Fraction         & 0.00 & 0.00 & 0 & 0\\
 & PID Systematics      & 0.00 & 0.01 & 0 & 0\\
 & Lineshapes           & 0.01 & 0.02 & 4 & 6\\
 & Fit Bias             & 0.01 & 0.00 & 1 & 1\\
 & Continuum DP         & 0.01 & 0.03 & 6 & 4\\
 & {\bf Total}          & {\bf 0.04} & {\bf 0.06} & {\bf 16} & {\bf 12}\\                                   
\hline\hline
\end{tabular}}
\end{center}
\end{table*}

\begin{table*}[h!]
\begin{center}
\caption{\label{tab:systTable_rho} Systematic uncertainties associated with the $\Bz\to\rho^-\Kp$ and non-resonant isobar parameters summarized in~\tabref{fourfitsolutions} under Sol. I. Uncertainties in the phases are in degrees.}
\setlength{\tabcolsep}{.9pc}
{\begin{tabular}{cccccc}
\hline\hline
Amplitude  & &  $\sigma (a)$ &  $\sigma (\overline{a})$   &  $\sigma (\Phi)~[^\circ]$ &  $\sigma (\overline{\Phi})~[^\circ]$ \\
\hline
\multirow{9}{*}{$\rho(770)^- K^+$}
 & Isobar Model          & 0.06 &fixed&fixed& fixed\\
 & \B Backgrounds         & 0.01 &fixed&fixed& fixed\\
 & PDF Shape Parameters  & 0.02 &fixed&fixed& fixed\\
 & SCF Fraction          & 0.00 &fixed&fixed& fixed\\
 & PID Systematics       & 0.01 &fixed&fixed& fixed\\
 & Lineshapes           & 0.01 &fixed&fixed& fixed\\
 & Fit Bias              & 0.01 &fixed&fixed& fixed\\
 & Continuum DP          & 0.01 &fixed&fixed& fixed\\
 & {\bf Total}           & {\bf 0.07} & {\bf fixed} & {\bf fixed} & {\bf fixed}\\
                   
\hline
\multirow{9}{*}{$\rho(1450)^- K^+$}
 & Isobar Model         & 0.04 & 0.03 & 25 & 5\\
 & \B Backgrounds        & 0.03 & 0.04 & 1 & 1\\
 & PDF Shape Parameters & 0.02 & 0.05 & 1 & 2\\
 & SCF Fraction         & 0.00 & 0.01 & 0 & 1\\
 & PID Systematics      & 0.00 & 0.02 & 1 & 0\\
 & Lineshapes           & 0.03 & 0.02 & 5 & 5\\
 & Fit Bias             & 0.02 & 0.01 & 2 & 4\\
 & Continuum DP         & 0.00 & 0.02 & 1 & 3\\
 & {\bf Total}          & {\bf 0.07} & {\bf 0.08} & {\bf 26} & {\bf 9}\\
                               
\hline
\multirow{9}{*}{$\rho(1700)^- K^+$}
 & Isobar Model         & 0.01 & 0.02 & 16 & 12\\
 & \B Backgrounds        & 0.04 & 0.02 & 2 & 3\\
 & PDF Shape Parameters & 0.02 & 0.02 & 3 & 0\\
 & SCF Fraction         & 0.01 & 0.00 & 0 & 1\\
 & PID Systematics      & 0.00 & 0.01 & 1 & 0\\
 & Line Shapes          & 0.04 & 0.05 & 10 & 9\\
 & Fit Bias             & 0.02 & 0.05 & 4 & 1\\
 & Continuum DP         & 0.00 & 0.01 & 0 & 4\\
 & {\bf Total}          & {\bf 0.06} & {\bf 0.08} & {\bf 20} & {\bf 16}\\                                 
\hline
\multirow{9}{*}{NR}
 & Isobar Model         & 0.01 & 0.00 & 13 & 1\\
 & \B Backgrounds        & 0.01 & 0.01 & 1 & 1\\
 & PDF Shape Parameters & 0.04 & 0.03 & 1 & 3\\
 & SCF Fraction         & 0.00 & 0.00 & 0 & 1\\
 & PID Systematics      & 0.00 & 0.01 & 0 & 1\\
 & Lineshapes           & 0.02 & 0.01 & 8 & 4\\
 & Fit Bias             & 0.05 & 0.01 & 1 & 2\\
 & Continuum DP         & 0.02 & 0.02 & 1 & 3\\
 & {\bf Total}          & {\bf 0.07} & {\bf 0.04} & {\bf 15} & {\bf 6}\\
\hline\hline
\end{tabular}}
\end{center}
\end{table*}

\begin{table*}[h]
\begin{center}
\caption{\label{tab:IF} The \CP~averaged interference fractions, $I_{kl}$ among the intermediate decay amplitudes expressed as a percentage of the total charmless decay amplitude. The interference fractions are calculated using the isobar amplitudes given in~\tabref{fourfitsolutions} as solution I.}
\setlength{\tabcolsep}{.5pc}
{\begin{tabular}{l|cccccccc}
\hline\hline
                      & $\rho(770)^- K^+$ & $\rho(1450)^- K^+$ & $\rho(1700)^- K^+$ & $\Kstar(892)^+\pi^-$ & $(K\pi)^{*+}_0\pi^-$ & $\Kstar(892)^0\pi^0$ & $(K\pi)^{*0}_0\pi^0$ & NR \\
\hline
$\rho(770)^- K^+$     & 17.61     & 7.22       & 0.88       & 0.47      & -1.49      & 0.50      & -0.78      & 0.00  \\
$\rho(1450)^- K^+$    &           & 6.34       & -1.71      & 0.60      & 0.65       & 0.42      & 0.97       & 0.00  \\ 
$\rho(1700)^- K^+$    &           &            & 1.68       & 0.22      & -0.72      & 0.23      & -0.28      & 0.00  \\
$\Kstar(892)^+\pi^-$  &           &            &            & 7.05      & 0.00       & -0.05     & -0.10      & 0.00  \\ 
$(K\pi)^{*+}_0\pi^-$  &           &            &            &           & 30.30      & -0.08     & 0.34       & -0.08 \\
$\Kstar(892)^0\pi^0$  &           &            &            &           &            & 5.87      & 0.00       & 0.00  \\ 
$(K\pi)^{*0}_0\pi^0$  &           &            &            &           &            &           & 15.29      & 1.16  \\
NR                   &           &            &            &           &            &           &            & 7.49  \\
\hline\hline  
\end{tabular}}
\end{center}
\end{table*}
\begin{table*}[tbh]
\begin{center}
\caption{\label{tab:Corr_Amp_I} Correlation coefficients among the floated isobar parameters for $\B$ decays.}
\setlength{\tabcolsep}{.5pc}
{\begin{tabular}{ll|cc|cc|cc}
\hline\hline
&  & \multicolumn{2}{c|}{$(K\pi)^{*+}_0\pi^-$} & \multicolumn{2}{c|}{$(K\pi)^{*0}_0\pi^0$} & \multicolumn{2}{c}{$\Kstar(892)^+\pi^-$} \\
&  & $a$ & $\Phi$ & $a$ & $\Phi$ & $a$ & $\Phi$ \\       
\hline
\multirow{2}{*}{$(K\pi)^{*+}_0\pi^-$}  & $a$    & 1.00  & -0.07  & 0.65  & -0.01  & 0.51  & -0.06 \\    
                     & $\Phi$  & -0.07  & 1.00  & 0.17  & 0.56  & 0.03  & 0.94   \\ 
\hline 
\multirow{2}{*}{$(K\pi)^{*0}_0\pi^0$}  & $a$     & 0.65  & 0.17  & 1.00  & 0.17  & 0.52  & 0.15  \\ 
                     & $\Phi$   & -0.01  & 0.56  & 0.17  & 1.00  & -0.01  & 0.51 \\  
\hline 
\multirow{2}{*}{$\Kstar(892)^+\pi^-$} & $a$     & 0.51  & 0.03  & 0.52  & -0.01  & 1.00  & 0.03 \\ 
                     & $\Phi$  & -0.06  & 0.94  & 0.15  & 0.51  & 0.03  & 1.00  \\  
\hline 
\multirow{2}{*}{$\Kstar(892)^0\pi^0$}  & $a$     & 0.56  & 0.01  & 0.36  & -0.04  & 0.42  & 0.00  \\
                     & $\Phi$   & 0.00  & 0.51  & 0.17  & 0.86  & -0.01  & 0.46  \\ 
\hline  
\multirow{2}{*}{NR}                   & $a$      & 0.43  & -0.38  & 0.15  & 0.02  & 0.37  & -0.36 \\           
                     & $\Phi$   & 0.07  & 0.63  & 0.18  & 0.68  & -0.05  & 0.53 \\   
\hline 
\multirow{2}{*}{$\rho(1450)^- K^+$}   & $a$    & 0.31  & -0.19  & 0.14  & -0.13  & 0.20  & -0.20 \\
                     & $\Phi$ & 0.06  & 0.52  & 0.11  & 0.58  & 0.04  & 0.46  \\  
\hline 
\multirow{2}{*}{$\rho(1700)^- K^+$}   & $a$      & 0.23  & -0.28  & 0.06  & -0.20  & 0.14  & -0.27 \\
                     & $\Phi$   & -0.12  & 0.62  & 0.10  & 0.56  & -0.03  & 0.56 \\  
\hline 
$\rho(770)^- K^+$    & $a$    & 0.56  & 0.17  & 0.56  & 0.20  & 0.45  & 0.16  \\
\hline\hline  
\end{tabular}}
\end{center}
\end{table*}

\begin{table*}[tbh]
\begin{center}
\caption{\label{tab:Corr_Amp_II} Correlation coefficients among the floated isobar parameters for $\B$ decays. }
\setlength{\tabcolsep}{.5pc}
{\begin{tabular}{ll|cc|cc|cc|cc|c}
\hline\hline
&  &  \multicolumn{2}{c|}{$K^{*0}(892)\pi^0$}  & \multicolumn{2}{c|}{NR} & \multicolumn{2}{c|}{$\rho(1450)^- K^+$} & \multicolumn{2}{c|}{$\rho(1700)^- K^+$} &$\rho(770)^- K^+$\\
&  &  $a$ & $\Phi$ & $a$ & $\Phi$ & $a$ & $\Phi$ & $a$ & $\Phi$ & $a$ \\ 
\hline
\multirow{2}{*}{$(K\pi)^{*+}_0\pi^-$}  & $a$    & 0.56  & 0.00  & 0.43  & 0.07  & 0.31  & 0.06  & 0.23  & -0.12  & 0.56  \\           
                     & $\Phi$ & 0.01  & 0.51  & -0.38  & 0.63  & -0.19  & 0.52  & -0.28  & 0.62  & 0.17 \\
\hline
\multirow{2}{*}{$(K\pi)^{*0}_0\pi^0$}  & $a$    & 0.36  & 0.17  & 0.15  & 0.18  & 0.14  & 0.11  & 0.06  & 0.10  & 0.56  \\
                     & $\Phi$ & -0.04  & 0.86  & 0.02  & 0.68  & -0.13  & 0.58  & -0.20  & 0.56  & 0.20  \\
\hline
\multirow{2}{*}{$\Kstar(892)^+\pi^-$}    & $a$    & 0.42  & -0.01  & 0.37  & -0.05  & 0.20  & 0.04  & 0.14  & -0.03  & 0.45 \\
                     & $\Phi$ & 0.00  & 0.46  & -0.36  & 0.53  & -0.20  & 0.46  & -0.27  & 0.56  & 0.16 \\
\hline
\multirow{2}{*}{$\Kstar(892)^0\pi^0$}    & $a$    & 1.00  & -0.09  & 0.25  & 0.09  & 0.27  & 0.18  & 0.24  & -0.10  & 0.42 \\
                     & $\Phi$ & -0.09  & 1.00  & 0.00  & 0.65  & -0.15  & 0.54  & -0.14  & 0.50  & 0.20 \\
\hline
\multirow{2}{*}{NR}                   & $a$    & 0.25  & 0.00  & 1.00  & -0.26  & 0.24  & -0.05  & 0.21  & -0.23  & 0.27 \\               
                     & $\Phi$ & 0.09  & 0.65  & -0.26  & 1.00  & -0.16  & 0.71  & -0.22  & 0.63  & 0.24 \\
\hline
\multirow{2}{*}{$\rho(1450)^- K^+$}   & $a$    & 0.27  & -0.15  & 0.24  & -0.16  & 1.00  & 0.02  & 0.70  & -0.67  & -0.10 \\
                     & $\Phi$ & 0.18  & 0.54  & -0.05  & 0.71  & 0.02  & 1.00  & 0.16  & 0.46  & 0.36 \\
\hline
\multirow{2}{*}{$\rho(1700)^- K^+$}   & $a$    & 0.24  & -0.14  & 0.21  & -0.22  & 0.70  & 0.16  & 1.00  & -0.61  & -0.01 \\
                     & $\Phi$ & -0.10  & 0.50  & -0.23  & 0.63  & -0.67  & 0.46  & -0.61  & 1.00  &  0.27 \\
\hline
$\rho(770)^- K^+$    & $a$    & 0.42  & 0.20  & 0.27  & 0.24  & -0.10  & 0.36  & -0.01  & 0.27  & 1.00 \\
\hline\hline
\end{tabular}}
\end{center}
\end{table*}

\begin{table*}[tbh]
\begin{center}
\caption{\label{tab:Corr_Ampbar_I} Correlation coefficients among the floated isobar parameters for $\Bb$ decays. }
\setlength{\tabcolsep}{.5pc}
{\begin{tabular}{ll|cc|cc|cc}
\hline\hline
&  & \multicolumn{2}{c|}{$(K\pi)^{*-}_0\pip$} & \multicolumn{2}{c|}{$(K\pi)^{*0}_0\pi^0$}  & \multicolumn{2}{c}{$\Kstar(892)^-\pip$}  \\
&   & $\overline{a}$ & $\overline{\Phi}$ & $\overline{a}$ & $\overline{\Phi}$ & $\overline{a}$ & $\overline{\Phi}$ \\       
\hline
\multirow{2}{*}{$(K\pi)^{*-}_0\pip$}   & $\overline{a}$       & 1.00  & 0.09  & -0.44  & 0.05  & -0.39  & 0.10  \\ 
                     & $\overline{\Phi}$    & 0.09  & 1.00  & 0.01  & 0.35  & 0.12  & 0.86   \\  
\hline
\multirow{2}{*}{$(K\pi)^{*0}_0\pi^0$}  & $\overline{a}$     & -0.44  & 0.01  & 1.00  & 0.47  & 0.34  & 0.02 \\ 
                     & $\overline{\Phi}$    & 0.05  & 0.35  & 0.47  & 1.00  & 0.02  & 0.28   \\  
\hline
\multirow{2}{*}{$\Kstar(892)^-\pip$}   & $\overline{a}$       & -0.39  & 0.12  & 0.34  & 0.02  & 1.00  & 0.14 \\  
                     & $\overline{\Phi}$    & 0.10  & 0.86  & 0.02  & 0.28  & 0.14  & 1.00   \\  
\hline
\multirow{2}{*}{$\Kstarb(892)^0\piz$}  & $\overline{a}$       & -0.49  & 0.01  & 0.24  & 0.03  & 0.31  & -0.01 \\ 
                     & $\overline{\Phi}$  & 0.07  & 0.34  & 0.34  & 0.80  & -0.02  & 0.28 \\   
\hline
\multirow{2}{*}{NR}                   & $\overline{a}$       & -0.49  & -0.25  & 0.25  & 0.02  & 0.30  & -0.23 \\
                     & $\overline{\Phi}$    & -0.09  & 0.33  & -0.11  & 0.45  & -0.08  & 0.18 \\ 
\hline
\multirow{2}{*}{$\rho(1450)^+\Km$}    & $\overline{a}$    & -0.63  & -0.19  & 0.41  & -0.17  & 0.35  & -0.18 \\
                     & $\overline{\Phi}$  & 0.02  & 0.48  & 0.02  & 0.50  & 0.02  & 0.37  \\  
\hline 
\multirow{2}{*}{$\rho(1700)^+\Km$}    & $\overline{a}$      & -0.37  & -0.17  & 0.24  & -0.13  & 0.22  & -0.16 \\
                     & $\overline{\Phi}$   & 0.30  & 0.49  & -0.19  & 0.36  & -0.14  & 0.39 \\   
\hline\hline
\end{tabular}}
\end{center}
\end{table*}

\begin{table*}[tbh]
\begin{center}
\caption{\label{tab:Corr_Ampbar_II} Correlation coefficients among the floated isobar parameters for $\Bb$ decays.}
\setlength{\tabcolsep}{.5pc}
{\begin{tabular}{ll|cc|cc|cc|cc}
\hline\hline
&  &  \multicolumn{2}{c|}{$\Kstarb(892)^0\piz$} & \multicolumn{2}{c|}{NR} & \multicolumn{2}{c|}{$\rho(1450)^+\Km$}  & \multicolumn{2}{c}{$\rho(1700)^+\Km$} \\
&  &  $\overline{a}$ & $\overline{\Phi}$ & $\overline{a}$ & $\overline{\Phi}$ & $\overline{a}$ & $\overline{\Phi}$ & $\overline{a}$ & $\overline{\Phi}$ \\       
\hline
\multirow{2}{*}{$(K\pi)^{*-}_0\pip$}   & $\overline{a}$     & -0.49  & 0.07  & -0.49  & -0.09  & -0.63  & 0.02  & -0.37  & 0.30 \\
                     & $\overline{\Phi}$  & 0.01  & 0.34  & -0.25  & 0.33  & -0.19  & 0.48  & -0.17  & 0.49 \\
\hline
\multirow{2}{*}{$(K\pi)^{*0}_0\pi^0$}  & $\overline{a}$     & 0.24  & 0.34  & 0.25  & -0.11  & 0.41  & 0.02  & 0.24  & -0.19 \\
                     & $\overline{\Phi}$  & 0.03  & 0.80  & 0.02  & 0.45  & -0.17  & 0.50  & -0.13  & 0.36 \\
\hline
\multirow{2}{*}{$\Kstar(892)^-\pip$}   & $\overline{a}$     & 0.31  & -0.02  & 0.30  & -0.08  & 0.35  & 0.02  & 0.22  & -0.14 \\
                     & $\overline{\Phi}$  & -0.01  & 0.28  & -0.23  & 0.18  & -0.18  & 0.37  & -0.16  & 0.39 \\
\hline
\multirow{2}{*}{$\Kstarb(892)^0\piz$}  & $\overline{a}$     & 1.00  & -0.05  & 0.25  & 0.05  & 0.35  & 0.03  & 0.23  & -0.16 \\
                     & $\overline{\Phi}$  & -0.05  & 1.00  & -0.01  & 0.49  & -0.23  & 0.49  & -0.18  & 0.35 \\
\hline
\multirow{2}{*}{NR}                   & $\overline{a}$     & 0.25  & -0.01  & 1.00  & -0.05  & 0.31  & 0.01  & 0.23  & -0.20 \\
                     & $\overline{\Phi}$  & 0.05  & 0.49  & -0.05  & 1.00  & -0.21  & 0.57  & -0.23  & 0.45 \\
\hline
\multirow{2}{*}{$\rho(1450)^+\Km$}    & $\overline{a}$     & 0.35  & -0.23  & 0.31  & -0.21  & 1.00  & -0.24  & 0.57  & -0.72 \\
                     & $\overline{\Phi}$  & 0.03  & 0.49  & 0.01  & 0.57  & -0.24  & 1.00  & 0.13  & 0.47 \\
\hline
\multirow{2}{*}{$\rho(1700)^+\Km$}    & $\overline{a}$     & 0.23  & -0.18  & 0.23  & -0.23  & 0.57  & 0.13  & 1.00  & -0.49 \\
                     & $\overline{\Phi}$  & -0.16  & 0.35  & -0.20  & 0.45  & -0.72  & 0.47  & -0.49  & 1.00 \\
\hline\hline
\end{tabular}}
\end{center}
\end{table*}
 
\end{document}